\documentclass[twocolumn]{revtex4-1}
\usepackage{graphicx}
\usepackage{graphics}
\usepackage{subfigure}
\usepackage{slashbox}
\usepackage{amsmath}
\usepackage{brackets}
\usepackage{psfrag}
\usepackage{epsfig}
\usepackage{hyperref}
\usepackage{amssymb}
\begin{document}

\ProvideTextCommandDefault{\textonehalf}{${}^1\!/\!{}_2\ $}

\title{Sensitivity estimate of proposed searches for exotic spin-dependent interactions using polarized helium}

\author{P.-H.~Chu}
\email[Email address: ]{pchu@lanl.gov}
\author{Y.~J.~Kim}
\email[Email address: ]{youngjin@lanl.gov}
\author{I.~Savukov}
\affiliation{Los Alamos National Laboratory, Los Alamos, New Mexico 87545, USA}
\date{\today}

\begin{abstract}
We investigate the sensitivities of searches for exotic spin-dependent interactions between the polarized nuclear spins of $^3$He and the particles of unpolarized or polarized solid-state masses using the frequency method and the resonance method. In the frequency method, the spin-dependent interactions act as an effective static magnetic field, causing the frequency shift to the spin precession of $^{3}$He. In the resonance method, proposed by Arvanitaki and Geraci [Phys. Rev. Lett. 113, 161801 (2014)] for the significant improvement of the experimental sensitivities on the spin-dependent interactions, the mass movement is modulated at the Larmor frequency of $^3$He. This results in the modulating spin-dependent interactions inducing an effective oscillatory magnetic field, which can tilt the $^3$He spins, similarly as an oscillatory magnetic field in nuclear magnetic resonance. We estimate the sensitivities of the searches using a room-temperature $^3$He target for its extremely long relaxation time. New limits on the coupling strengths of the spin-dependent interactions can be set in the interaction length range below $10^{-1}$ m.
\end{abstract}

\maketitle
Since the discovery of intrinsic spin~\cite{Commins:2012}, exotic spin-dependent interactions between fermions have been of interest. Moody and Wilczek~\cite{Moody:1984} first considered some types of exotic interactions between polarized spins and unpolarized or polarized particles through new spin-0 boson exchange. Later, Dobrescu and Mocioiu~\cite{Dobrescu:2006au} extended this idea by including the operators dependent on the relative velocity between two interacting particles in the non-relativistic limit through new spin-1 boson exchange. Recently, these exotic spin-dependent interactions have attracted people's attention because they are observables of new spin-0 or spin-1 bosons, which may solve several mysteries in fundamental physics. For example, the axion as a spin-0 boson was introduced in the theory to explain the lack of charge-parity (CP) violation in the strong interaction~\cite{Peccei:1977} and the cold dark matter~\cite{Duffy:2009ig}. Several theoretical concepts including string theory~\cite{Arvanitaki:2010}, hierarchy problem~\cite{Graham:2015}, dark energy~\cite{Flambaum:2009}, unparticles~\cite{Georgi:2007}, dark photons~\cite{Appelquist:2003,Dobrescu:2005,Ackerman:2009} also predict the existence of such new bosons. A review article describing the recent theoretical progress in this field can be found in Ref.~\cite{Safronova:2017xyt}.

There are fifteen possible exotic spin-dependent interactions described in Ref.~\cite{Dobrescu:2006au}, which have been revisited in a convenient format~\cite{Leslie:2014mua,Fadeev:2018rfl}. The contact terms have been also studied for the superficial singularity~\cite{Fadeev:2018rfl,Fadeev:2019jzi}. In this paper we still use the fifteen interaction formats adopting the numbering scheme in Ref.~\cite{Dobrescu:2006au} since we will not consider the contact term. In a system of two particles (particle 1, 2 are fermions like electrons, neutrons, protons, etc.) with spin 1 ($\hat{\sigma}_{1}$) and 2 ($\hat{\sigma}_{2}$), and mass 1 ($m_1$) and 2 ($m_2$) respectively, their relative distance and relative velocity are $\vec{r}$ and $\vec{v}$. We can group the spin-dependent interactions between these two particles as static spin-dependent interactions, spin-velocity-dependent interactions and spin-velocity-velocity-dependent interactions. The group one includes the interactions:
\begin{eqnarray}
V_{2} &=& f_{2}\frac{\hbar c}{4\pi}\left(\hat{\sigma}_1\cdot \hat{\sigma}_2\right)\left(\frac{1}{ r}\right)e^{-r/\lambda},\label{eq:v2}
\\
V_{3} &=& f_{3}\frac{\hbar^{3}}{4\pi m_{1}m_{2} c}\left[(\hat{\sigma}_1\cdot\hat{\sigma}_2)\left(\frac{1}{\lambda r^2}+\frac{1}{r^{3}}\right)\right.\notag\\
&-& \left.(\hat{\sigma}_1\cdot\hat{r})(\hat{\sigma}_2\cdot\hat{r})\left(\frac{1}{\lambda^2 r}+\frac{3}{\lambda r^2}+\frac{3}{r^3}\right)\right]e^{-r/\lambda},
\label{eq:v3}\\
V_{9+10} &=&  f_{9+10}\frac{\hbar^{2}}{8\pi m_{1}}(\hat{\sigma}_{1}\cdot\hat{r})\left(\frac{1}{\lambda r}+\frac{1}{r^{2}}\right)e^{-r/\lambda},
\label{eq:v910}\\
V_{11} &= & -f_{11}\frac{\hbar^2}{4\pi m_\mu}[(\hat{\sigma}_1\times\hat{\sigma}_2)\cdot\hat{r}]\left(\frac{1}{\lambda r}+\frac{1}{r^2}\right)e^{-r/\lambda}.
\label{eq:v11}
\end{eqnarray}
The group two includes the interactions:
\begin{eqnarray}
&V_{4+5}&=  -f_{4+5}\frac{\hbar^{2}}{8\pi m_{1}c}\left[\hat{\sigma}_{1}\cdot(\vec{v}\times\hat{r})\right]\nonumber\\
&\times&\left(\frac{1}{\lambda r}+\frac{1}{r^{2}}\right)e^{-r/\lambda},\label{eq:v45}
 \\
&V_{12+13}& =  f_{12+13}\frac{\hbar}{8\pi}(\hat{\sigma}_{i}\cdot\vec{v})\left(\frac{1}{r}\right)e^{-r/\lambda},
\label{eq:v1213}\\
&V_{6+7}&= -f_{{6+7}}\frac{\hbar^{2}}{4\pi m_{\mu}c}\nonumber\\
&\times&\left[(\hat{\sigma}_1\cdot\vec{v})(\hat{\sigma}_2\cdot\hat{r})\right]\left(\frac{1}{\lambda r}+\frac{1}{r^{2}}\right)e^{-r/\lambda},\label{eq:v67}
 \\
 &V_{14}&  =  f_{14}\frac{\hbar}{4\pi}[(\hat{\sigma}_1\times\hat{\sigma}_2)\cdot\vec{v}]\left(\frac{1}{r}\right)e^{-r/\lambda},
\label{eq:14}
\\
&V_{15}&=-f_{15}\frac{\hbar^3}{8\pi m_1 m_2 c^2 }\nonumber\\
&\times&\{[\hat{\sigma}_1\cdot(\vec{v}\times\hat{r})](\hat{\sigma}_2\cdot\hat{r})+(\hat{\sigma}_1\cdot\hat{r})[\hat{\sigma}_2\cdot(\vec{v}\times\hat{r})]\}\nonumber\\
&\times&\left(\frac{1}{\lambda^2 r}+\frac{3}{\lambda r^2}+\frac{3}{r^3}\right)e^{-r/\lambda},
\label{eq:v15}
\end{eqnarray}	
and the group three includes the interactions:
\begin{eqnarray}
V_{8} & = & f_8\frac{\hbar}{4\pi c}(\hat{\sigma}_1\cdot\vec{v})(\hat{\sigma}_2\cdot\vec{v})\left(\frac{1}{r}\right)e^{-r/\lambda},
\label{eq:v8}\\
V_{16} & = & -f_{16}\frac{\hbar^2}{8\pi m_\mu c^2 }\nonumber\\
&\times&\{[\hat{\sigma}_1\cdot(\vec{v}\times\hat{r})](\hat{\sigma}_2\cdot\vec{v})+(\hat{\sigma}_1\cdot\vec{v})[\hat{\sigma}_2\cdot(\vec{v}\times\hat{r})]\}\nonumber\\
&\times&\left(\frac{1}{\lambda r}+\frac{1}{ r^2}\right)e^{-r/\lambda}
\label{eq:v16}
\end{eqnarray}	
where $m_\mu$ is the reduced mass of $m_1$ and $m_2$, and $\lambda$ is the interaction length. $f_{i}$'s are the coupling strengths that we measure, which can be the combination of scalar, pseudoscalar, vector and axial-vector coupling~\cite{Leslie:2014mua, Fadeev:2018rfl}. All spin-dependent interactions have the potential form as $\hat{\sigma}_1\cdot\vec{A}$, which is similar to the Zeeman interaction term of a spin with a magnetic field, $\hat{\sigma}_1\cdot\vec{B}$~\cite{Chu:2016}, indicating $\vec{A}$ can affect a spin like an ordinary magnetic field.

Several experimental methods on various spin-dependent interactions over a broad interaction length range have been conducted, including spectroscopy, torsion-pendulum, magnetometry, parity nonconservation and electric dipole moment experiments~\cite{Safronova:2017xyt,Fadeev:2018rfl}. However, most experimental searches are still related to static spin-dependent interactions including $V_{2}, V_{3}, V_{9+10}$ and $V_{11}$. For the group two and the group three, Yan and Snow used neutron beams to study the spin-velocity-dependent interaction ($V_{12+13}$)~\cite{Yan:2012wk} and later
Yan {\it et al.} used the relaxation of polarized $^{3}$He to explore the same interaction in different interaction range~\cite{Yan:2015}. Adelberger and Wagner also set new constraints by combining different experimental limits~\cite{Adelberger:2013}. 
Piegsa and Pignol used Ramsey’s technique of separated oscillatory fields with a cold neutron beam to investigate $V_{4+5}$~\cite{Piegsa:2012}. Haddock {\it et al.} applied a slow neutron polarimeter that passed transversely polarized slow neutrons by unpolarized slabs of material to set a new constraint of $V_{4+5}$~\cite{Haddock_2018}. Chu {\it et al.} proposed to use spin exchange relaxation-free (SERF) magnetometers to search for exotic spin-dependent interactions~\cite{Chu:2016} and later set new limits on $V_{4+5}$~\cite{Kim:2017yen} and $V_{12+13}$~\cite{Kim:2019sry} between polarized electrons and unpolarized nucleons. Hunter {\it et al.} first applied polarized geoelectrons to search for long-range spin-spin interactions~\cite{Hunter:2013} and later expanded the idea to the velocity-dependent spin-dependent interactions~\cite{Hunter:2014}. Ji {\it et al.} proposed to use K-$^{3}$He spin-exchange-relaxation-free (SERF) comagnetometers with SmCo$_5$ spin sources~\cite{Ji:2017} to search for exotic spin-dependent interactions and later Ji {\it el al} used K-Rb SERF comagnetometers with SmCo$_5$ spin sources to set new limits on spin-spin-velocity-dependent interactions~\cite{Ji:2018}. Leslie {\it et al.} proposed to search exotic spin-dependent interactions with rare earth iron garnet test masses (dysprosium iron garnet, DyIG)~\cite{Leslie:2014mua} while the paramagnetic insulator, gadolinium
gallium garnet (GGG), also has potential for spin-dependent interactions~\cite{Chu:2015tha}. 

\begin{figure}[t]
\centering
\includegraphics[width=0.5\textwidth]{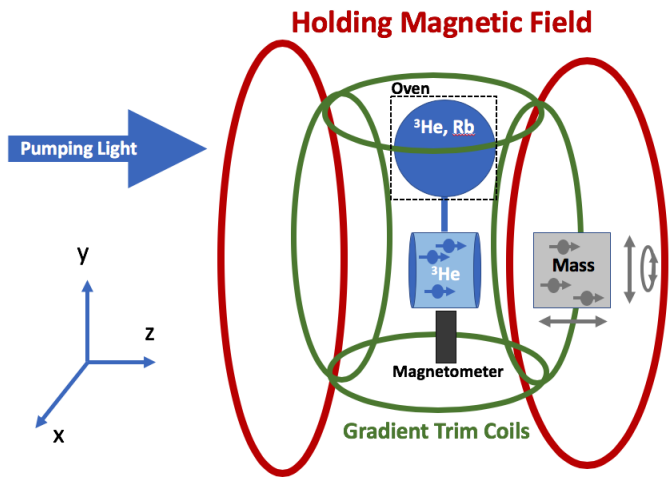}
\caption{The schematic of the experiment. The pumping chamber in an oven contains $^3$He and Rb atoms. $^{3}$He spins can be polarized by the pumping light through the spin-exchange with Rb electron spins. The target chamber is at the center of the holding magnetic field. Gradient trim coils are used to improve the transverse relaxation time $T_2$. The mass is close to the target chamber. Using different  configurations of mass motions and positions, the setup can be sensitive to different spin-dependent interactions. A sensitive magnetometer will be used to detect the magnetization of $^3$He. }
\label{fig:schematic}
\end{figure}

In this paper, we consider two methods, the frequency method~\cite{Chu:2012cf, Chu:2016} and the resonance method~\cite{Arvanitaki:2014dfa} using spin-exchange optical pumping (SEOP) polarized $^{3}$He targets~\cite{Gentile_2017} with unpolarized or polarized mass. The polarization of the $^3$He spins using SEOP can be close to $p \approx 1$~\cite{Gentile_2017}.  The polarized $^3$He spins can precess at the Larmor frequency $\omega_N = \gamma_3 B_0$ in a static magnetic field $\vec{B}_0$ where the gyromagnetic ratio $\gamma_3 = (2\pi)\times 32.4$ MHz/T. The induced magnetization of $^3$He can be measured using a sensitive magnetometer with the sensitivity from $10^{-12}$ to  $10^{-15}$ T. Optical pumping magnetometers (OPM) can in general reach the sensitivity of $10^{-15}$ T~\cite{Budker_2007}. But a reduction factor should be considered due to the geometry and the interference between $^3$He and OPM. So we assume a conservative sensitivity of the magnetometer at $10^{-12}$ T. A typical SEOP $^{3}$He cell has a double-cell configuration as shown in Fig.~\ref{fig:schematic}~\cite{Gentile_2017}. The top cell is a spherical pumping chamber filled with $^3$He and Rb atoms. The pumping chamber is usually in an oven in order to create Rb vapor gas for optical pumping. The bottom target chamber contains only $^3$He because of temperature difference between two chambers. The pressure of $^3$He is at the order of 1 atm. The cell wall thickness is about 1 mm while a thinner wall about the order of 250 $\mu$m is feasible~\cite{Chu:2012cf}. The target chamber is at the center of the $\vec{B}_0$ with gradient trim coils in order to optimize the relaxation time~\cite{Rosenberry:2001}. 

We plan to use BGO (bismuth germanate, $\text{Bi}_{4}\text{Ge}_{3}\text{O}_{12}$) as the unpolarized mass for its high nuclear density ($4.29\times 10^{30} \text{m}^{-3}$) and small magnetic effects~\cite{Tullney:2013}. DyIG (dysprosium iron garnet, $\text{Dy}^{3+}_{3}\text{Fe}^{3+}_{2}\text{Fe}^{3+}_{3} \text{O}_{12}$) is proposed as the polarized mass which has shown the property of near-zero magnetization at the critical temperature around 220--240K and the spin density is about $10^{26}\text{m}^{-3}$~\cite{Leslie:2014mua}.  In the following estimation, the BGO mass is assumed to be a cube of 2 cm used in Ref.~\cite{Kim:2017yen} and Ref.~\cite{Kim:2019sry} and the DyIG mass is assumed to be a cylinder with the radius of 0.4 cm and the length 0.2 cm~\cite{Leslie:2014mua}.   

\begin{table}[t]
\caption{Geometry of experiment for each interactions for the frequency method. $\hat{\sigma}_1$, $\hat{\sigma}_2$, $\vec{v}$, $\vec{B}_{\text{eff}}$, $\delta r$ for different interactions.}
\begin{center}
\begin{tabular}{ c | c | c| c | c| c | c}
\hline
interaction &  $\hat{\sigma}_1$ & $\hat{\sigma}_2$ & $\vec{v}$ & position & $\vec{B}_{\text{eff}}$ & $\delta r (mm)$\\
\hline
\hline
$V_{2}$ & $\hat{z}$ &  $\hat{z}$ & $\hat{z}$  & $z$ & $\hat{z}$ &1\\
\hline
$V_{3}$ & $\hat{z}$ &  $\hat{z}$ & $\hat{z}$  & $z$ & $\hat{z}$ &1\\
\hline
$V_{4+5}$ & $\hat{z}$ & 0 & $\hat{\phi}$ & $z$ & $\hat{z}$ & 1\\
\hline
$V_{6+7}$ & $\hat{z}$ &  $\hat{z}$ & $\hat{z}$  & $z$ & $\hat{z}$ &15\\
\hline
$V_{8}$ & $\hat{z}$ &  $\hat{z}$ & $\hat{z}$  & $z$ & $\hat{z}$ &15\\
\hline
$V_{9+10}$ & $\hat{z}$ & 0 & $\hat{z}$  & $z$ & $\hat{z}$ &1\\
\hline
$V_{11}$ & $\hat{x}$ &  $\hat{y}$ & $\hat{z}$  & $z$ & $\hat{x}$ &1\\
\hline
$V_{12+13}$ & $\hat{z}$ &  0 & $\hat{z}$  & $z$ & $\hat{z}$ &6\\
\hline
$V_{14}$ & $\hat{x}$ &  $\hat{y}$ & $\hat{z}$  & $z$ & $\hat{x}$ &15\\
\hline
$V_{15}$ & $\hat{z}$ &  $\hat{z}$ & $\hat{\phi}$  & $z$ &$\hat{z}$ & 10\\
\hline
$V_{16}$ & $\hat{x}$ &  $\hat{y}$ & $\hat{y}$  & $z$ & $\hat{x}$ &10\\
\hline
\end{tabular}
\end{center}
\label{tab:experiment_freq}
\end{table}

In the frequency method, the spin-dependent interactions act as an effective static magnetic field $\vec{B}_\text{eff}$ shifting the precession frequency. The sensitivity estimation is based on two experiments ~\cite{Chu:2012cf,Tullney:2013}. Chu {\it et al.}~\cite{Chu:2012cf} applied the method of gradiometers, using two pickup coils to measure different parts of the $^{3}$He target, which have different effects from the spin-dependent interactions. In the environment without magnetic shielding, their frequency sensitivity was about $10^{-5}$ Hz, corresponding to $B_\text{eff}\sim 3\times 10^{-13}$ T for $^{3}$He. Tullney {\it et al.}~\cite{Tullney:2013} applied the Xe-$^{3}$He comagnetometer in the magnetically shielded room, measuring the precession of two atomic species, which have different effects from the spin-dependent interactions. They can reach the sensitivity of $10^{-9}$ Hz, corresponding to $B_\text{eff}\sim3\times 10^{-17}$ T for $^{3}$He. The practical sensitivity of $B_\text{eff}$ is probably between these two values if using a magnetically shielded room with gradient trim coils. Table~\ref{tab:experiment_freq} shows the experimental configurations using the frequency method for each interaction. The $\hat{\sigma}_{1}$ means the spin orientation of the $^{3}$He along the $\vec{B}_0$. The $\hat{\sigma}_{2}$ means the spin orientation of the polarized mass. The $\vec{v}$ is the velocity direction of the mass, where $\hat{\phi}$ means the rotation around the $\hat{z}$-axis, $\hat{z}$ means the movement along the $\hat{z}$-axis, vice versa. The position means the position of the mass relative to the $^{3}$He target chamber, where we assume the mass is always at the side to the $z$-axis. The direction of  $\vec{B}_{\text{eff}}$ is always along the $\vec{B}_0$ direction. $\delta r$ means the relative minimum distance from the mass to the $^{3}$He target chamber. $\delta r = 1$ mm means the unpolarized mass can touch the target chamber while $\delta = 5$ mm means the unpolarized mass needs to have a small distance for the velocity-dependent effect. $\delta r = 10$ mm means an additional thermal insulator thickness for the polarized mass to touch the target while $\delta r = 15$ mm means an additional distance for the velocity-dependent effect for the polarized mass. If the thermal insulator is not needed, for example, a vacuum system is applied, then $\delta r$ can be reduced. Although at the critical temperature the polarized mass DyIG has zero magnetization, the temperature fluctuation could induce additional magnetic field noise. Additional magnetic shields may be necessary to reduce the magnetic effects but increase $\delta r$.
\begin{table}[t]
\caption{Parameter values used in Eq.~\ref{eq:Mx}. }
\begin{center}
\begin{tabular}{ c | c | c}
\hline
Parameters & Symbol & Value\\
\hline
\hline
polarization & $p$ & 1 \\\hline 
spin density  of $^{3}$He & $n_s$ & 2.4$\times 10^{25}$ m$^{-3}$\\\hline
nuclear magnetic moment of $^{3}$He & $\mu_3$ & -1.07$\times 10^{-26}$ J/T \\\hline
gyromagnetic ratio of $^{3}$He& $\gamma_3$ & 2.03$\times 10^{8}$ Hz/T\\\hline
transverse relaxation time & $T_2$ & 1000 s or 53 h \\\hline
sensitivity of transverse magnetization & $\mu_0 M_x$ & $10^{-12}-10^{-14}$ T\\\hline
BGO nucleon density & & 4.29$\times 10^{30}$ m$^{-3}$\\
\hline
DyIG spin density & & 1$\times 10^{26}$ m$^{-3}$\\
\hline
\hline
\end{tabular}
\end{center}
\label{tab:parameters}
\end{table}
\begin{table}[t]
\caption{Geometry of experiment for each interactions for the resonance method. $\hat{\sigma}_1$, $\hat{\sigma}_2$ and $\vec{v}$, $\vec{B}_\text{eff}$, $\delta r$ for different interactions.}
\begin{center}
\begin{tabular}{ c | c | c| c | c| c | c}
\hline
interaction &  $\hat{\sigma}_1$ & $\hat{\sigma}_2$ & $\vec{v}$ & position & $\vec{B}_{\text{eff}}$ & $\delta r$ (mm)\\
\hline
\hline
$V_{2}$ & $\hat{x}$ &  $\hat{z}$ & $\hat{z}$  & $z$ & $\hat{z}$ &1\\
\hline
$V_{3}$ & $\hat{x}$ &  $\hat{z}$ & $\hat{z}$  & $z$ & $\hat{z}$ &1\\
\hline
$V_{4+5}$ & $\hat{x}$ & 0 & $\hat{\phi}$ & $z$ & $\hat{z}$ & 1\\
\hline
$V_{6+7}$ & $\hat{x}$ &  $\hat{z}$ & $\hat{z}$  & $z$ & $\hat{z}$ &15\\
\hline
$V_{8}$ & $\hat{x}$ &  $\hat{z}$ & $\hat{z}$  & $z$ & $\hat{z}$ &15\\
\hline
$V_{9+10}$ & $\hat{x}$ & 0 & $\hat{z}$  & $z$ & $\hat{z}$ & 1\\
\hline
$V_{11}$ & $\hat{z}$ & $\hat{z}$ & $\hat{y}$ &  $z$ & $\hat{x}$ &1\\
\hline
$V_{12+13}$ & $\hat{x}$ &  0 & $\hat{z}$  & $z$ & $\hat{z}$ & 6\\
\hline
$V_{14}$ & $\hat{z}$ &  $\hat{y}$ & $\hat{z}$  & $z$ & $\hat{x}$ &15\\
\hline
$V_{15}$ & $\hat{x}$ &  $\hat{z}$ & $\hat{\phi}$  & $z$ & $\hat{z}$ & 10\\
\hline
$V_{16}$ & $\hat{z}$ &  $\hat{y}$ & $\hat{y}$  & $z$ & $\hat{x}$ & 10 \\
\hline
\end{tabular}
\end{center}
\label{tab:experiment_resonance}
\end{table}

In the resonance method~\cite{Arvanitaki:2014dfa}, we consider the experiment in the room temperature environment in order to simplify the experimental apparatus as a pathfinder for the low temperature experiment. The mass movement is modulated at $\omega_N$ so that the spin-dependent interactions can induce an effective oscillatory magnetic field $\vec{B}_\text{eff}(t)\approx \vec{B}_{\text{eff}}\cos(\omega_N t)$. The $\vec{B}_{\text{eff}}(t)$ perpendicular to the $\vec{B}_0$ can rotate spins from the Bloch's equation~\cite{Bloch:1946, Chu:2015bta}. The $B_0$ should be less than 100 nT for the precession frequency less than 3 Hz, which is feasible for most motors. The gradiometer method can be considered, while the comagnetometer method unlikely works because of different resonance frequencies between the two atomic species.  The key point of the resonance method is that other magnetic field noise cannot rotate the spins as the linear oscillatory magnetic field. Therefore, the time-varying transverse magnetization $M_x$ of $^{3}$He~\cite{Bloch:1946, Arvanitaki:2014dfa} scales linearly in response to the small $B_{\text{eff}}$ until the measurement time $t\approx T_2$ as 
\begin{align}
    M_x(t) \approx \frac{1}{2}p n_s \mu_3\gamma_{3}B_{\text{eff}}T_2(e^{-t/T_1}-e^{-t/T_2})\cos(\omega_N t)
    \label{eq:Mx}
\end{align}
where $n_s = 2.4\times 10^{19} \text{cm}^{-3}$ is the spin density in the $^3$He target chamber which can be calculated using the ideal gas law $PV=n_sRT$ with the pressure $P$ = 1 atm, the volume $V$ = 1 cm$^3$ and the room temperature $T$ = 300 K, and $\mu_3 = -2.12\times\mu_N= -1.07\times 10^{-26}\text{J/T}$ is the nuclear magnetic moment of $^{3}$He, $T_1$ and $T_2$ are the longitudinal and transverse relaxation time ($T_1\gg T_2$).  The magnetization fluctuation of $^3$He~\cite{Arvanitaki:2014dfa} is $\sqrt{M_N^2} = \sqrt{\hbar\gamma_3 n_s \mu_3 T_2/2V}\approx 2\times10^{-15}$ T$/\mu_0$, which is at the same level of the sensitivity of the optical pumping magnetometer. However, the background noise magnetic field in a magnetically shielded room is about $10^{-15}$ T/$\sqrt{\text{Hz}}$. For 1 second measurement, the noise level of $M_x$ is about $\sim10^{-14}$ T which can be calculated using Eq.~\ref{eq:Mx} with $B_\text{eff} = 10^{-15}$ T and $t = 1$, determining the sensitivity limit of $M_x$. Therefore, we estimate the sensitivity of $\mu_0M_x$ about $10^{-12}-10^{-14}$ T. If the amplitude of the traverse magnetization $M_x$ of $^3$He is $\sim 10^{-12}$ T, the worst sensitivity of the magnetometer, after 1000 s of measurement, this implies the  $B_\text{eff}$ upper limit is $\sim 3\times 10^{-17}$ T. We also list the scenario with the $M_x\sim 10^{-14}$ T, the background noise limit, and the extremely long $T_2 = 53$ h which has been achieved in Ref.~\cite{Tullney:2013}, giving the sensitivity of the $B_\text{eff} \sim 3\times 10^{-21}$ T.  The practical sensitivity is probably between $3\times 10^{-17}$ T and $3\times 10^{-21}$ T. The corresponding sensitivities of SDIs can be further improved by several cycles of measurement. The corresponding parameters for Eq.~\ref{eq:Mx} is summarized in Table~\ref{tab:parameters}.

\begin{figure*}[h]
\centering
\includegraphics[width=0.4\textwidth]{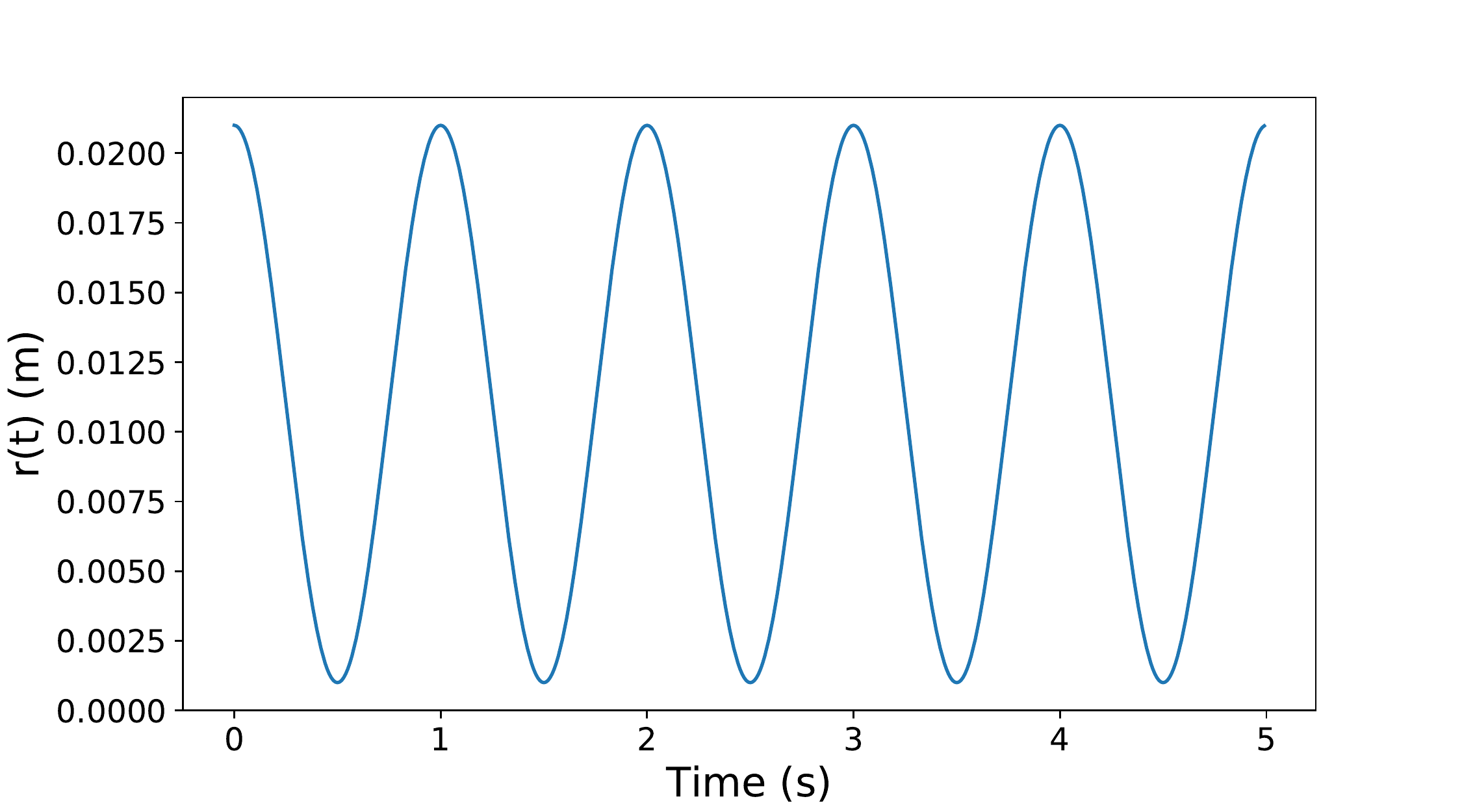}
\includegraphics[width=0.4\textwidth]{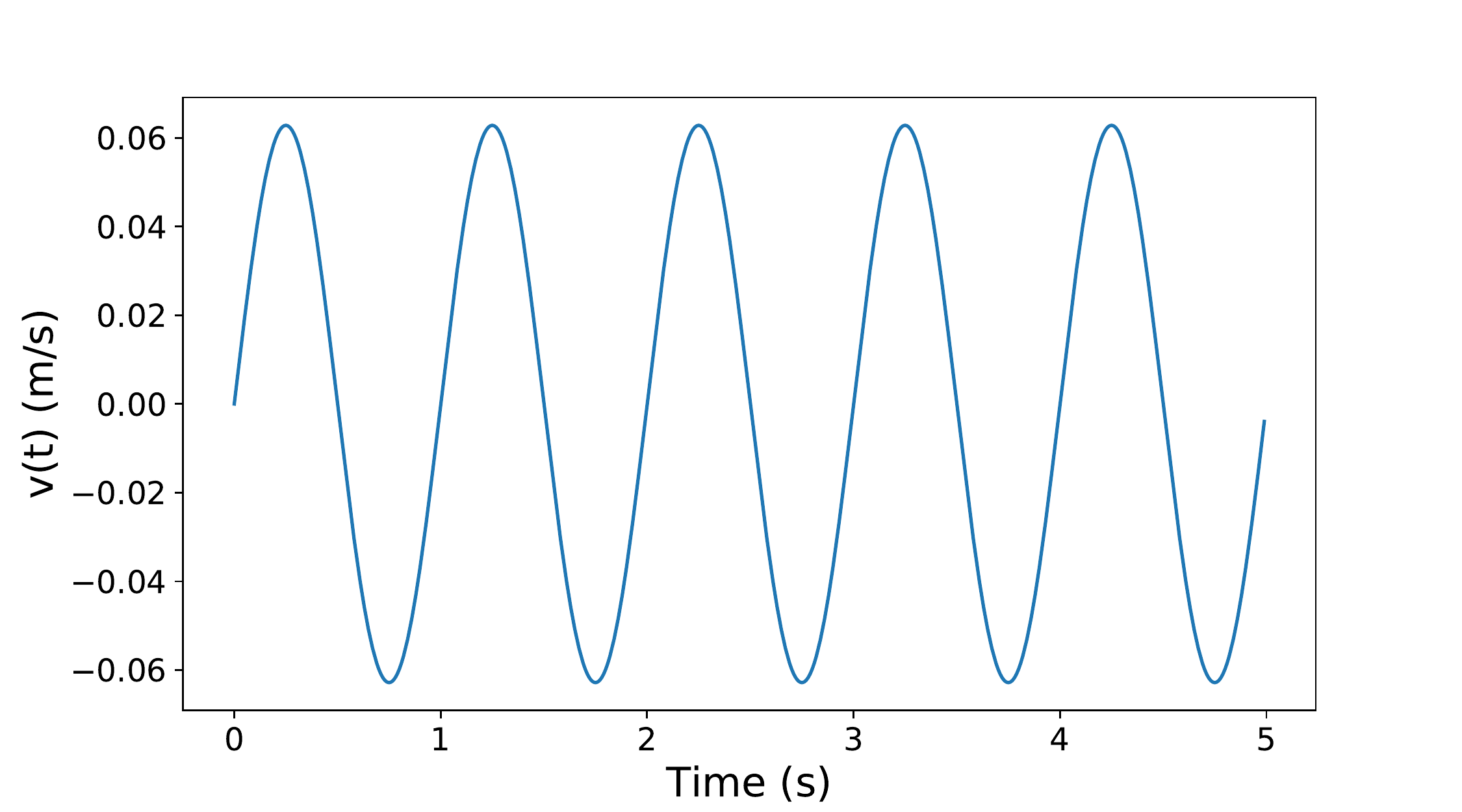}
\includegraphics[width=0.4\textwidth]{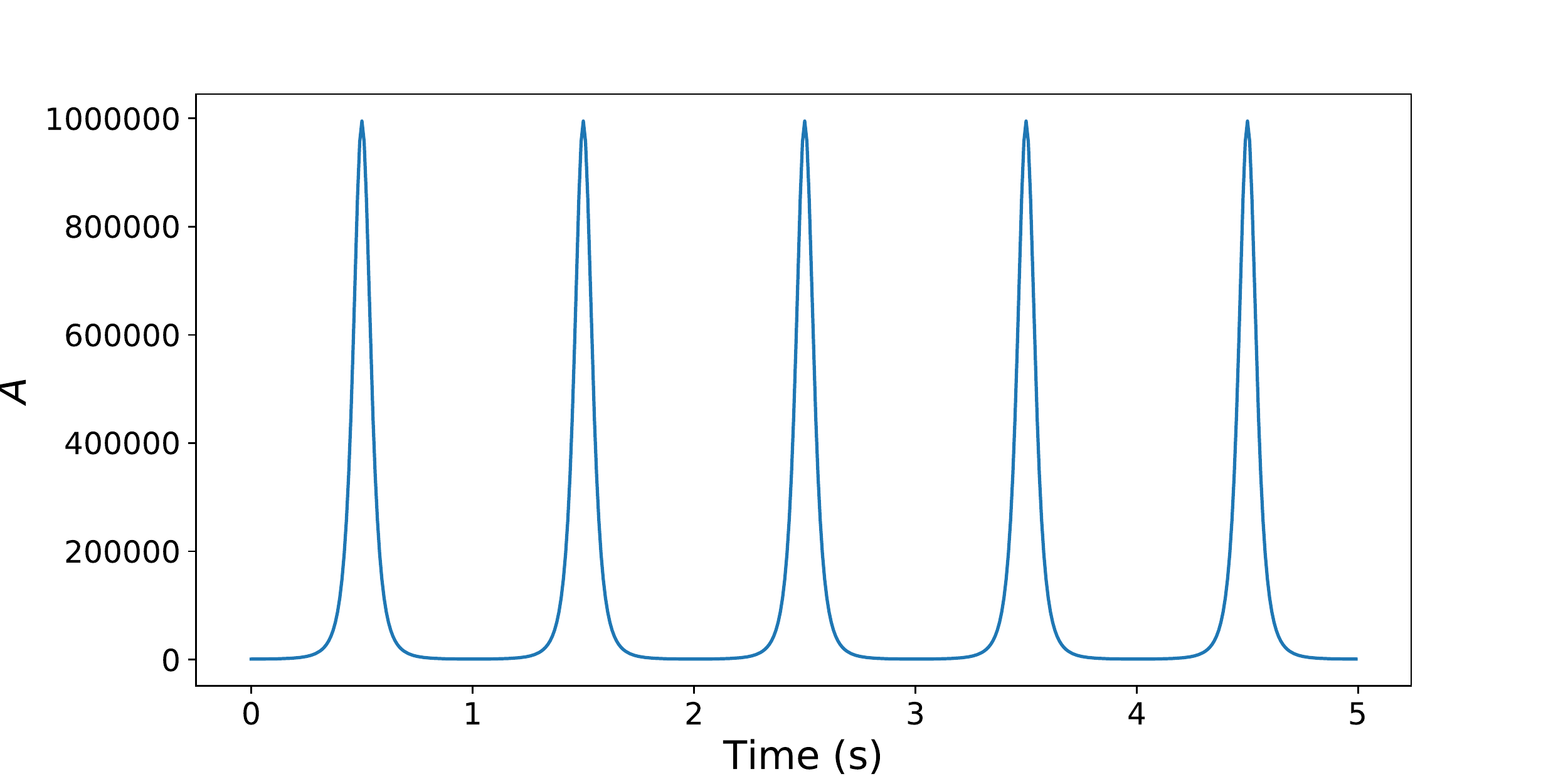}
\includegraphics[width=0.4\textwidth]{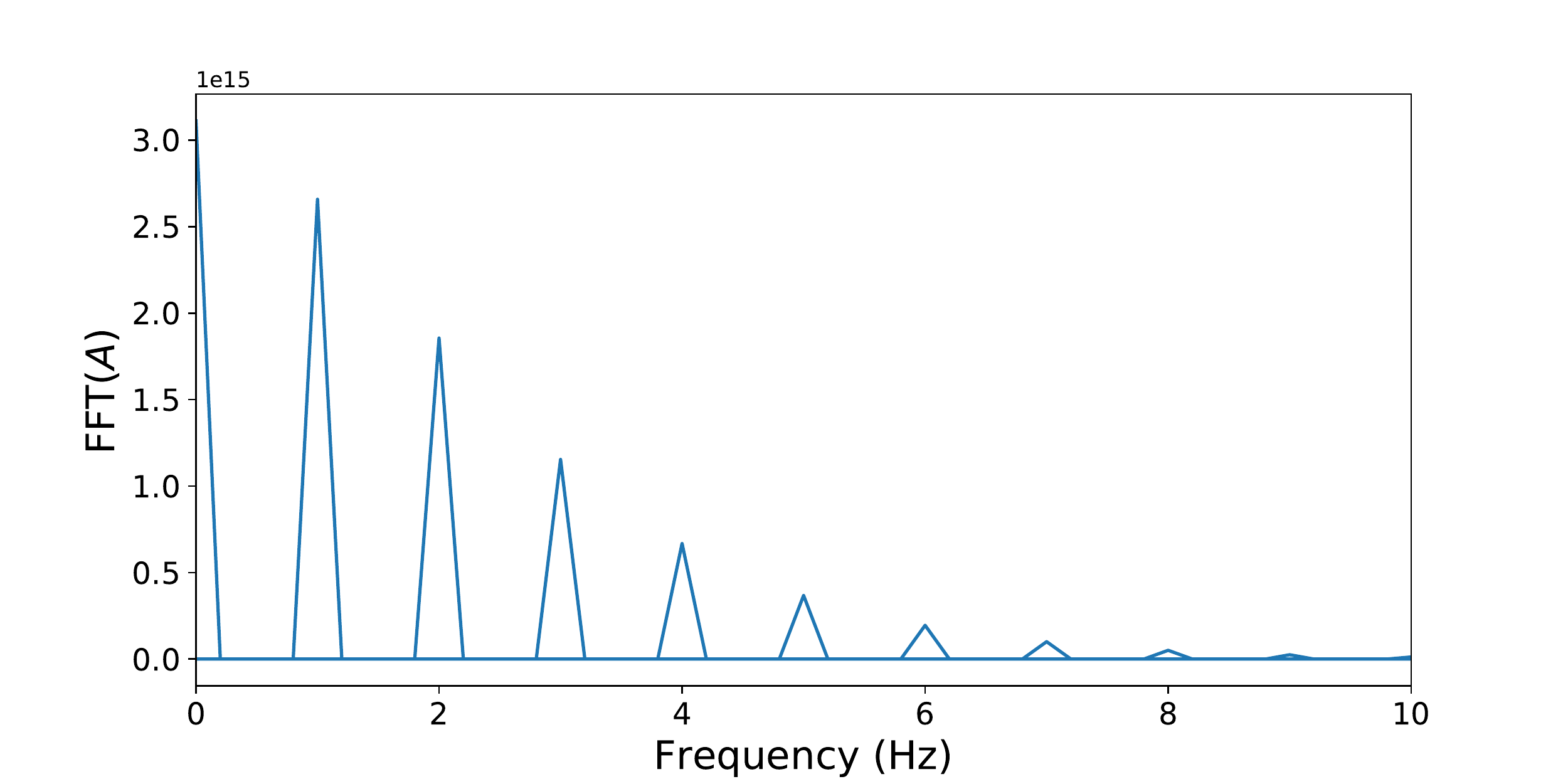}
\caption{The distance between the mass and the target is $r(t) = 0.01(1+\cos(\omega_N t))+0.001$ and the velocity is $v(t) = 0.01\omega_N \sin(\omega_N t)$ where $\omega_N = 2\pi f_N = 2\pi\times 1$ Hz. If $\lambda = 0.01$ m, the $B_{\text{eff}}$ due to $V_{9+10}$ is proportional to $A \equiv (\frac{1}{\lambda r}+\frac{1}{r^2})e^{-r/\lambda}$  which is roughly a function of $\sin(\omega_N t)$. The fast Fourier transform of $A$ is FFT($A$) showing the component of $f_N$. }
\label{fig:Beff_9}
\end{figure*}

\begin{figure*}[h]
\centering
\includegraphics[width=0.4\textwidth]{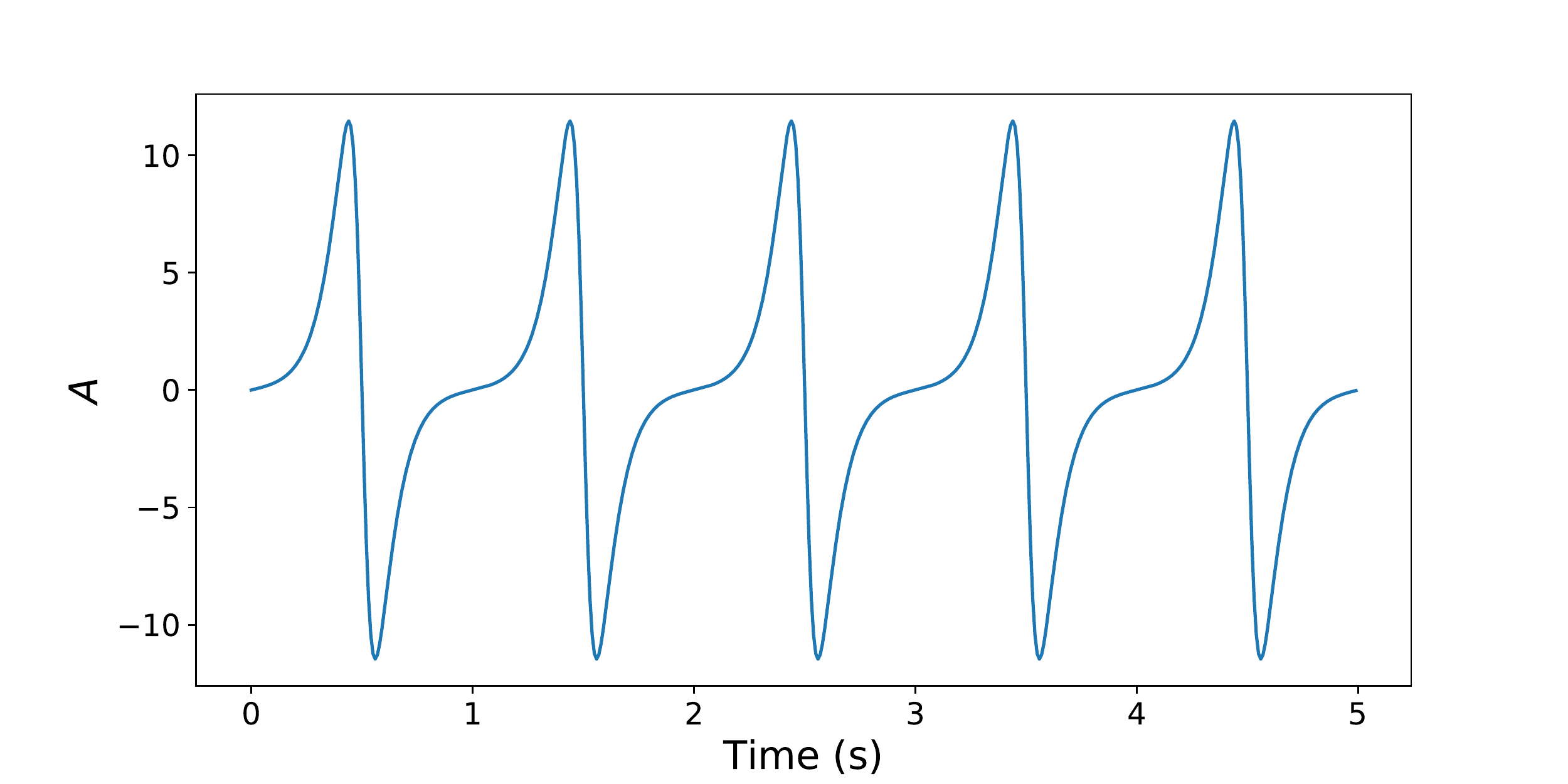}
\includegraphics[width=0.4\textwidth]{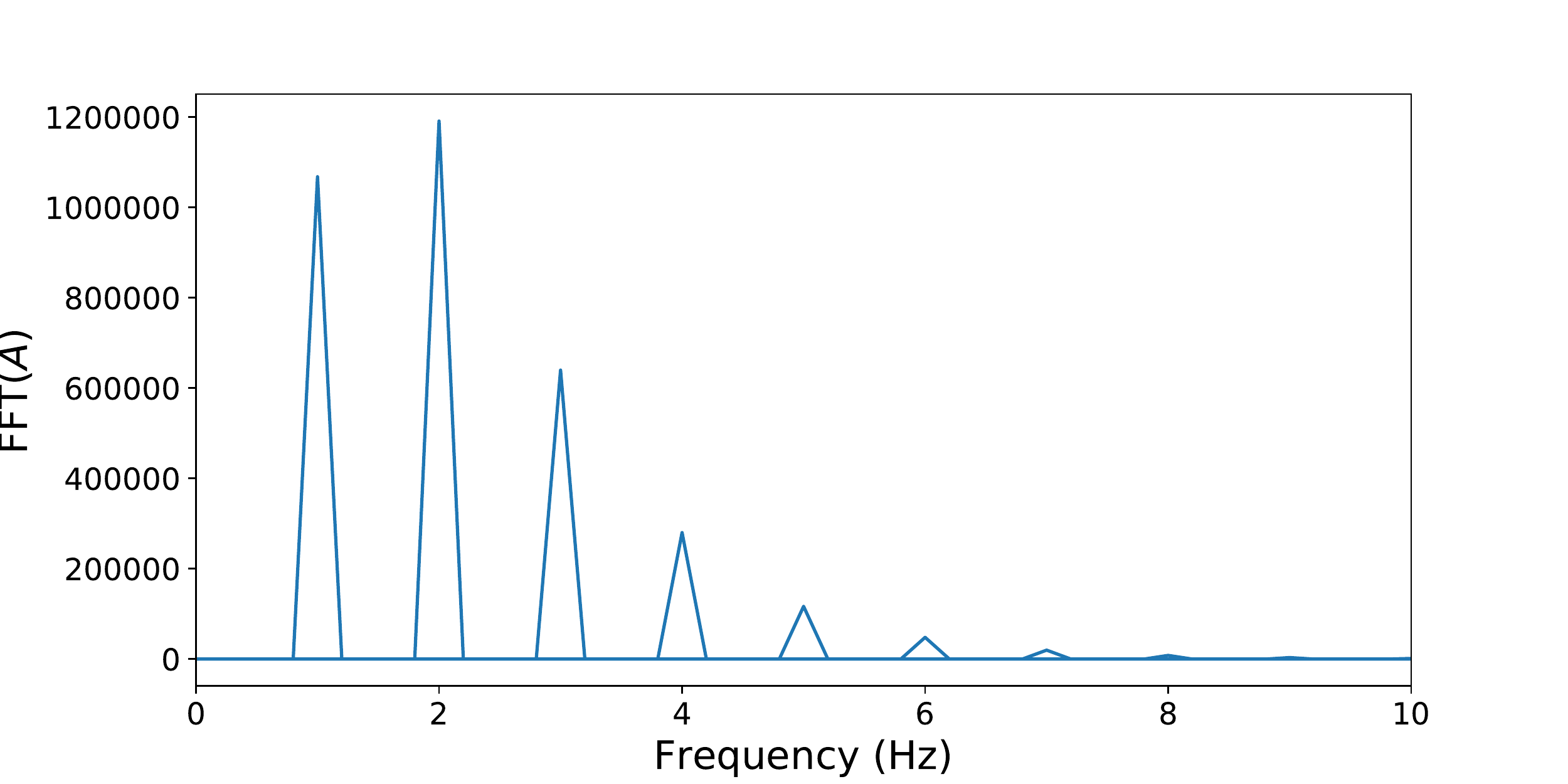}
\caption{If $\lambda = 0.01$ m, the  $B_{\text{eff}}$ due to $V_{12+13}$ is proportional to $A \equiv v(\frac{1}{ r})e^{-r/\lambda}$. The fast Fourier transform of $A$ is FFT($A$) showing the component of $f_N$. The peak of $f_N = 1$ Hz is relatively weaker.}
\label{fig:Beff_12}
\end{figure*}

\begin{figure*}[h]
\centering
\includegraphics[width=0.4\textwidth]{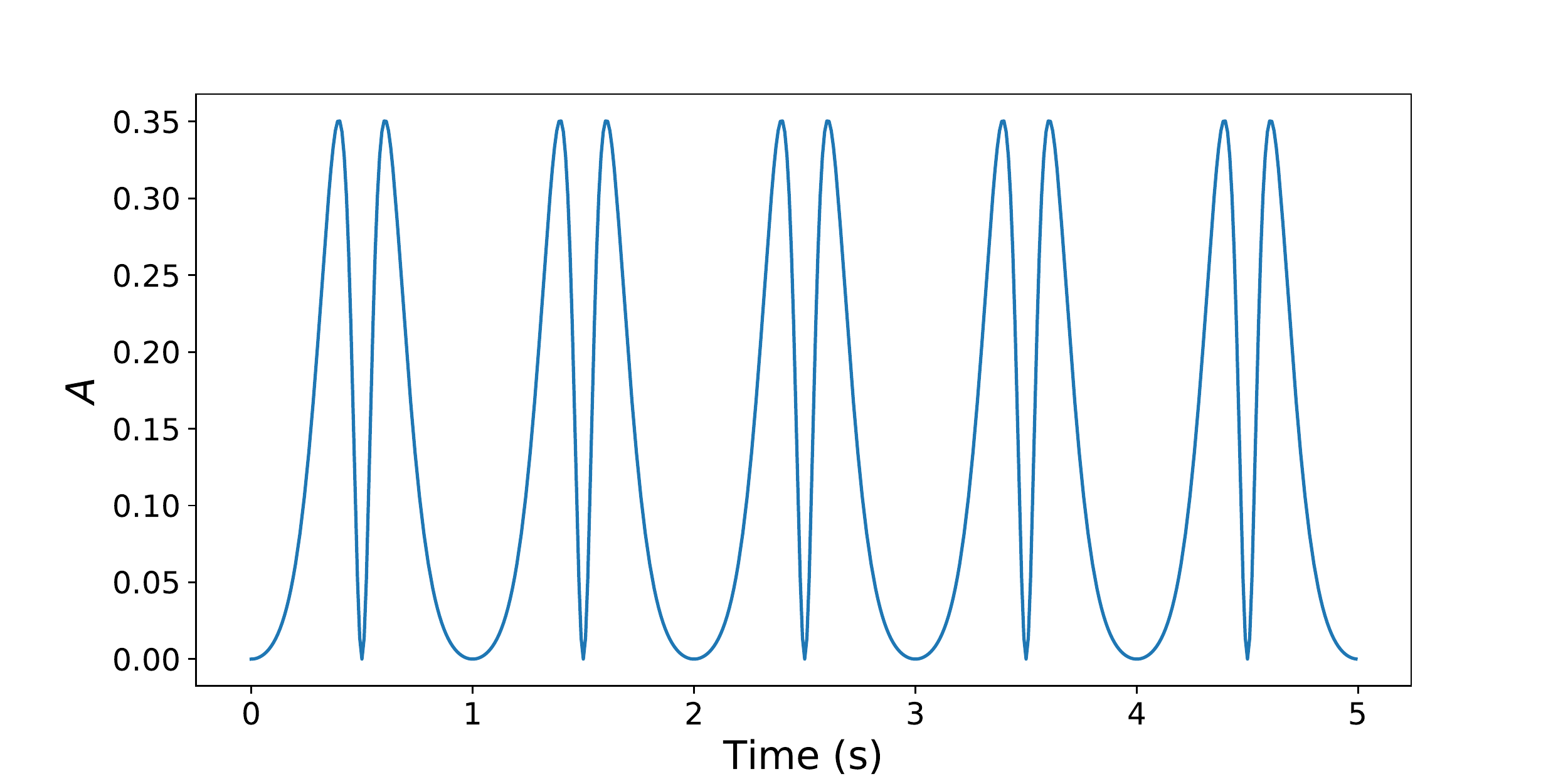}
\includegraphics[width=0.4\textwidth]{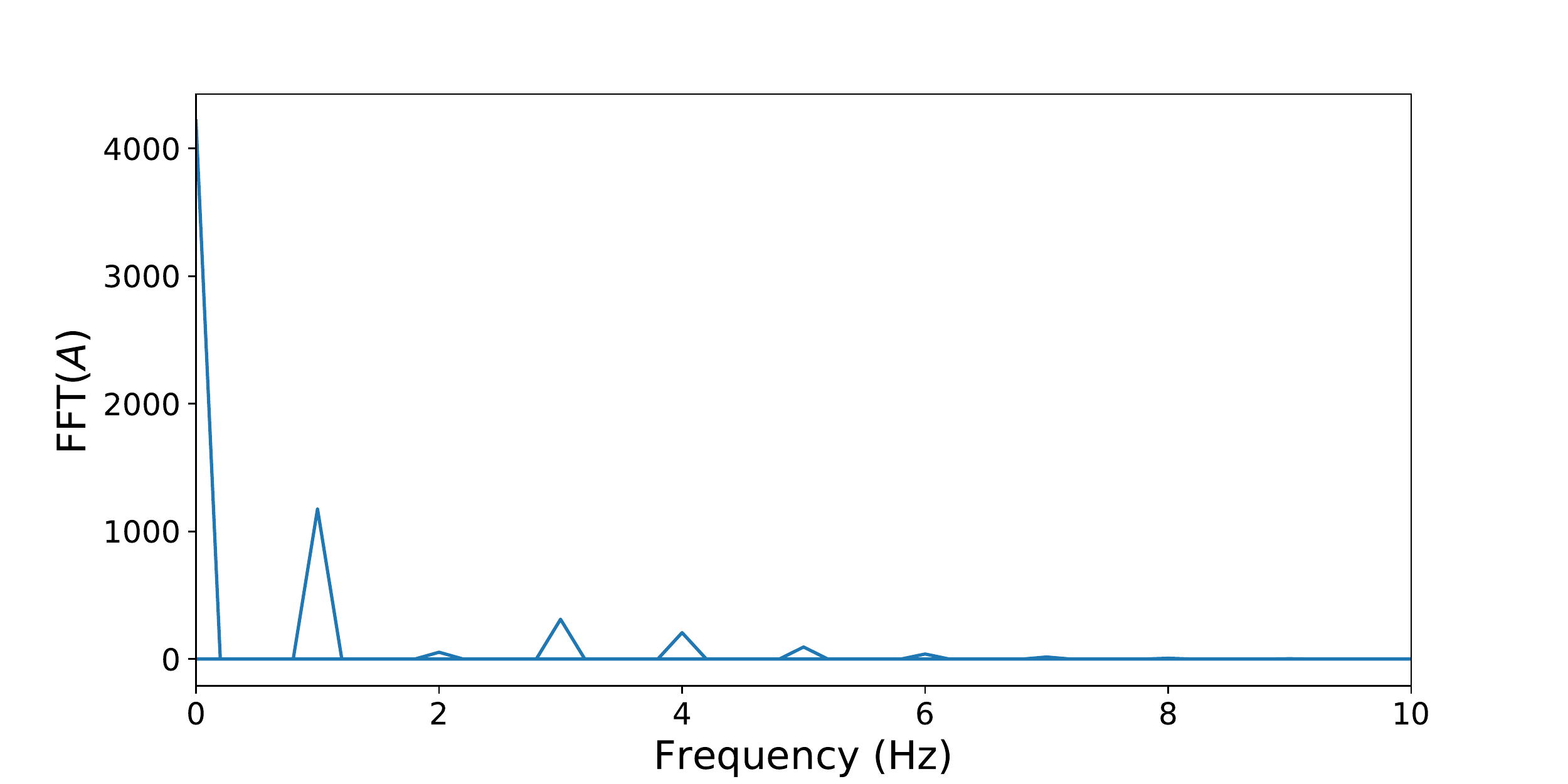}
\caption{If $\lambda = 0.01$ m, the  $B_{\text{eff}}$ due to $V_{8}$ is proportional to $A \equiv v^2(\frac{1}{ r})e^{-r/\lambda}$. The fast Fourier transform of $A$ is FFT($A$) showing the component of $f_N$. }
\label{fig:Beff_8}
\end{figure*}

Table~\ref{tab:experiment_resonance} shows the experimental configurations using the resonance method for each interaction. The definition of each parameter is the same as Table~\ref{tab:experiment_freq}. The  $\vec{B}_{\text{eff}}$ direction is always perpendicular to the $\vec{B}_0$ direction. For the static spin-dependent interaction such as $V_{2}$, $V_{3}$, $V_{9+10}$, and $V_{11}$, the $\vec{B}_\text{eff}$ can be modulated by the distance $r(t)=r(1+\cos{(\omega_N t)})+\delta r$ so that $B_{\text{eff}}(t)\approx B_{\text{eff}}(1+\cos(\omega_N t))$.  Figure~\ref{fig:Beff_9} shows the example of $V_{9+10}$. The $B_\text{eff}$ is proportional to $A\equiv (\frac{1}{\lambda r}+\frac{1}{r^2})e^{-r/\lambda}$, showing spikes with the frequency equal to $\omega_N$. The high-frequency components of $B_\text{eff}(t)$ larger than $\omega_N$ should be neglected, and a reduction factor should be considered when estimating the sensitivity to the $B_\text{eff}$ in Eq.~\ref{eq:Mx}. For the spin-velocity-dependent interactions such as $V_{4+5}$, $V_{12+13}$, $V_{6+7}$, $V_{14}$ and $V_{15}$, it is straightforward to modulate the velocity such as $v(t) = v\cos(\omega_N t)$ so that $B_{\text{eff}}(t)\approx B_{\text{eff}}\cos{(\omega_N t)}$. Figure~\ref{fig:Beff_12} shows the example of $V_{12+13}$ using the same modulation as in Fig.~\ref{fig:Beff_9}. The $B_\text{eff}$ is proportional to $A\equiv (\frac{v}{r})e^{-r/\lambda}$, showing spikes with the frequency equal to $\omega_N$. For the spin-velocity-velocity-dependent interactions such as $V_{8}$ and $V_{16}$, the $B_\text{eff}(t)$ at the resonance of $\omega_N$ can be still generated by the same modulation of the velocity as in Fig.~\ref{fig:Beff_9}.  Figure~\ref{fig:Beff_8} shows the example of $V_{8}$ and the $B_\text{eff}$ is proportional $A \equiv v^2(\frac{1}{ r})e^{-r/\lambda}$. However, the fast Fourier transform (FFT) of $A(t)$ implies that the reduction factor of $B_\text{eff}$ is larger so that the sensitivity for the spin-velocity-velocity-dependent interactions is much weakened.

The dominant systematic uncertainty in this proposal is the magnetic impurities buried in the mass. For the spin-velocity-dependent interactions, this uncertainty can be efficiently suppressed by reversing mass moving direction with the frequency method~\cite{Kim:2017yen, Kim:2019sry}. For other interactions and the resonance method, this systematic uncertainty could be mitigated with additional magnetic shields, however, which might affect the sensitivity of magnetometers~\cite{Lee_2008}. The geometry of the magnetometer and the magnetic shields should be optimized.  

\begin{figure*}[h]
\centering
\includegraphics[width=0.45\textwidth]{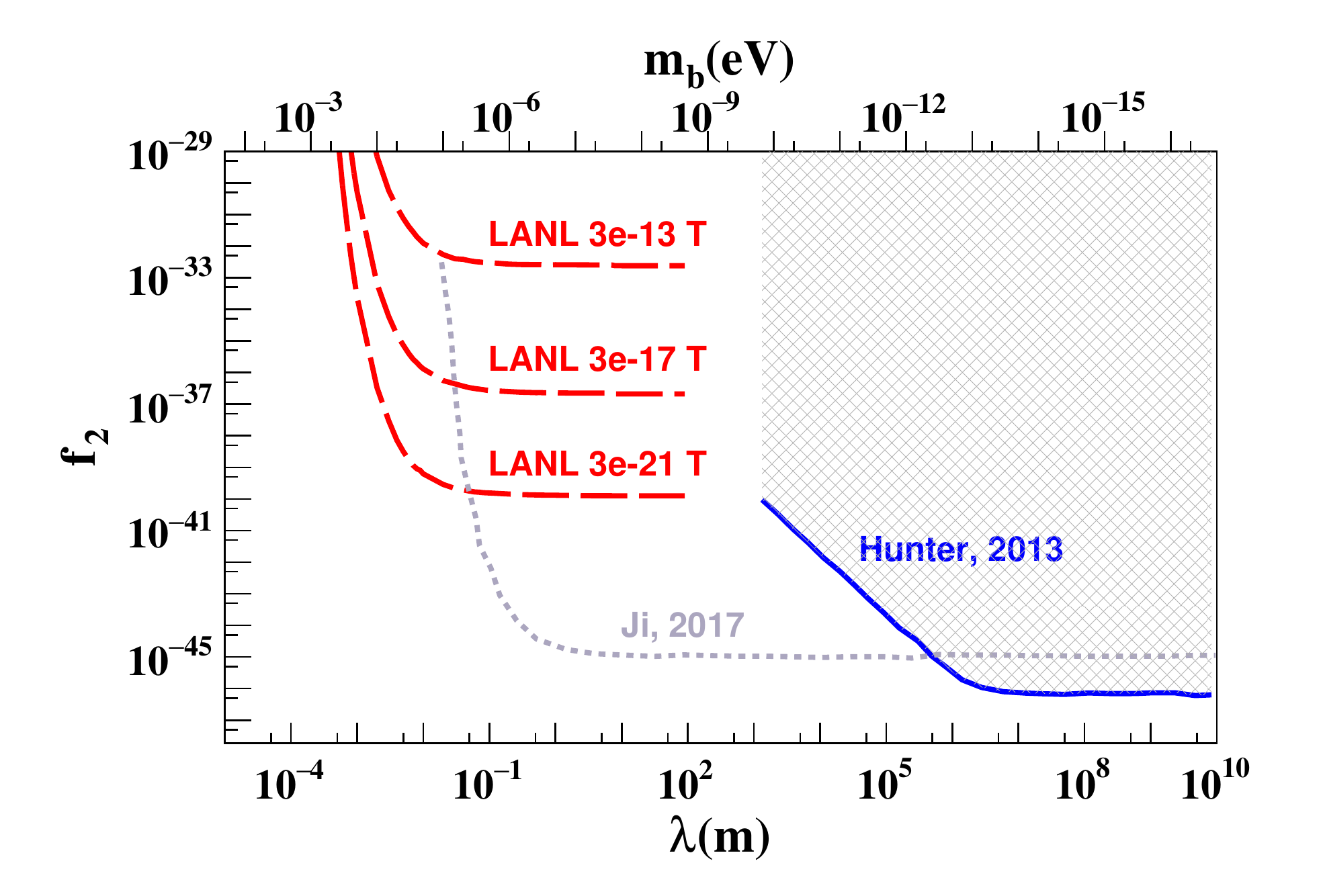}
\includegraphics[width=0.45\textwidth]{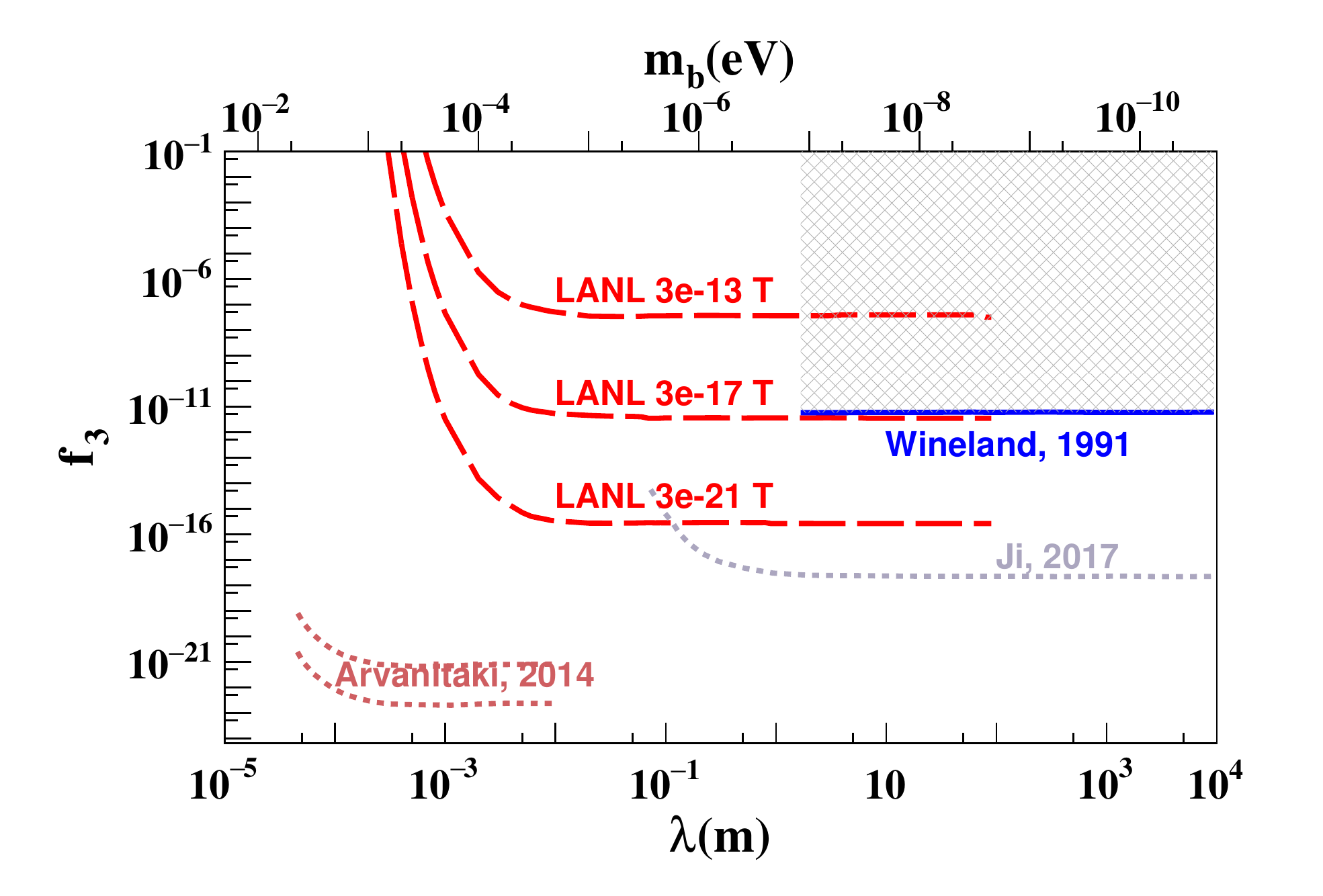}
\includegraphics[width=0.45\textwidth]{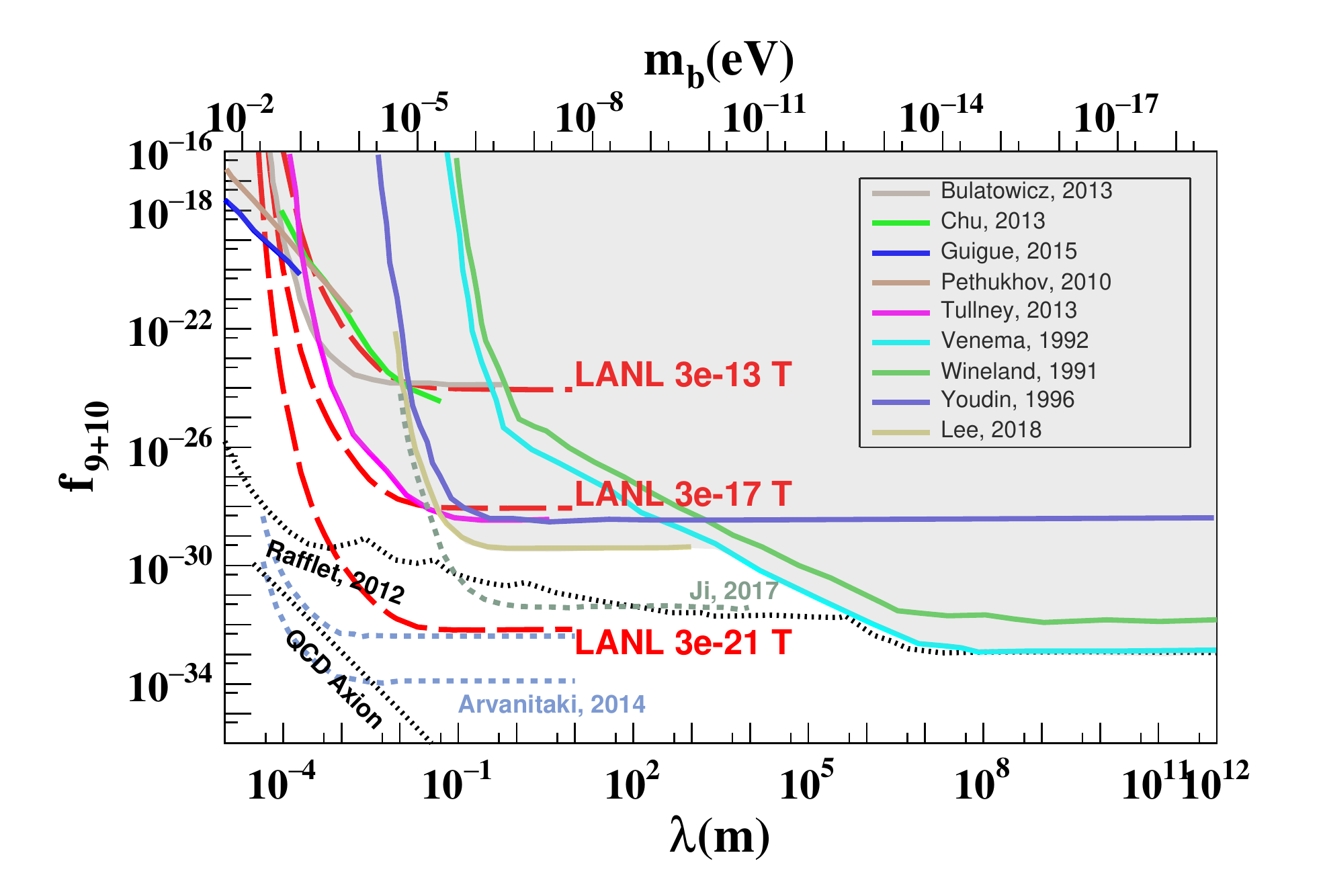}
\includegraphics[width=0.45\textwidth]{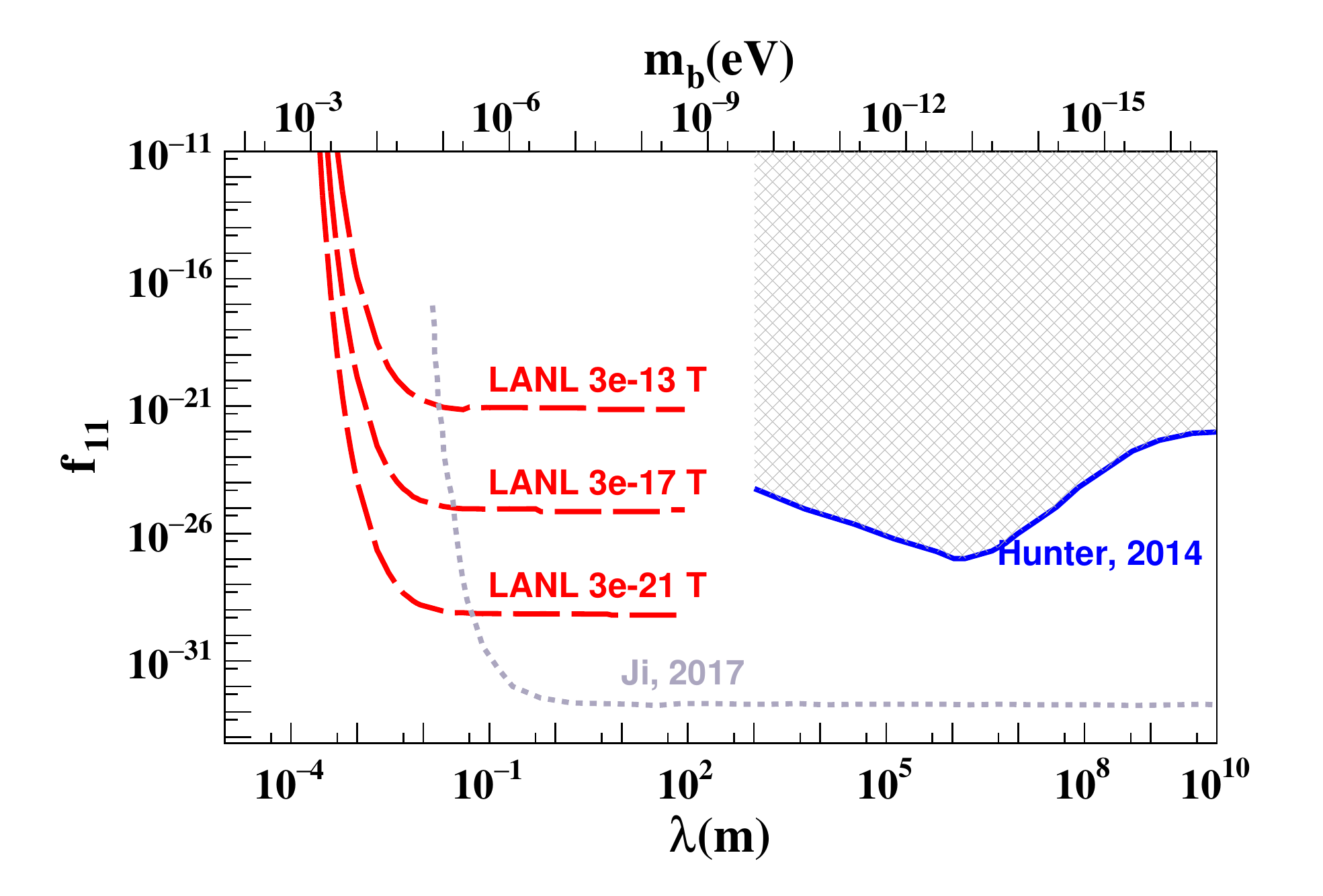}
\caption{The limits and the sensitivity estimation of 
$V_{2}$~\cite{Hunter:2013,Ji:2017}, $V_{3}$~\cite{Wineland:1991,Ji:2017,Arvanitaki:2014dfa},$V_{9+10}$~\cite{Wineland:1991,Venema:1992,Youdin:1996,Petukhov:2010,Chu:2012cf,Bulatowicz:2013,Tullney:2013,Guigue:2015fyt,Lee:2018vaq,Raffelt:2012sp,Ji:2017,Arvanitaki:2014dfa}, and $V_{11}$~\cite{Hunter:2013,Ji:2017}.}
\label{fig:copuling_0}
\end{figure*}

\begin{figure*}[h]
\centering
\includegraphics[width=0.45\textwidth]{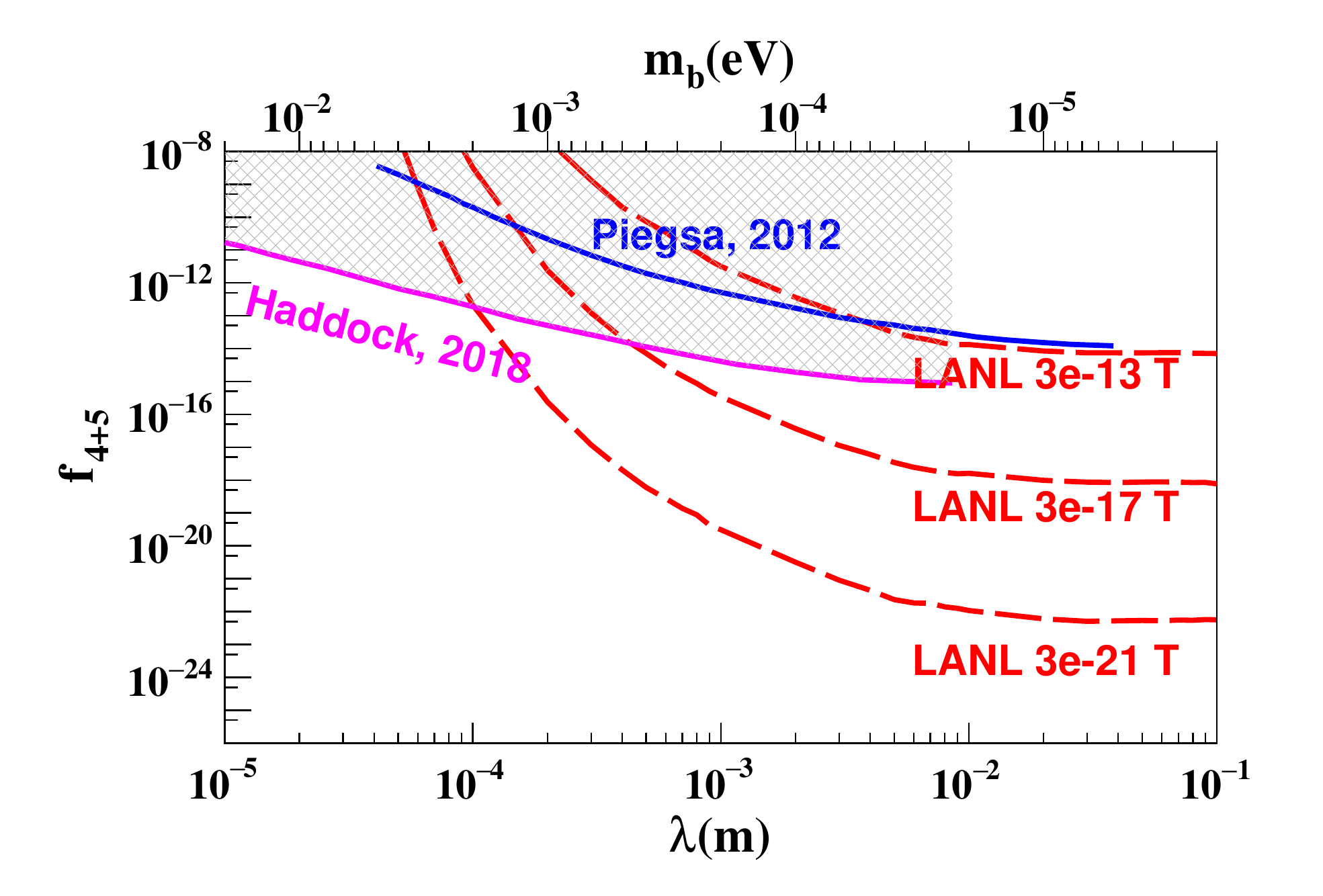}
\includegraphics[width=0.45\textwidth]{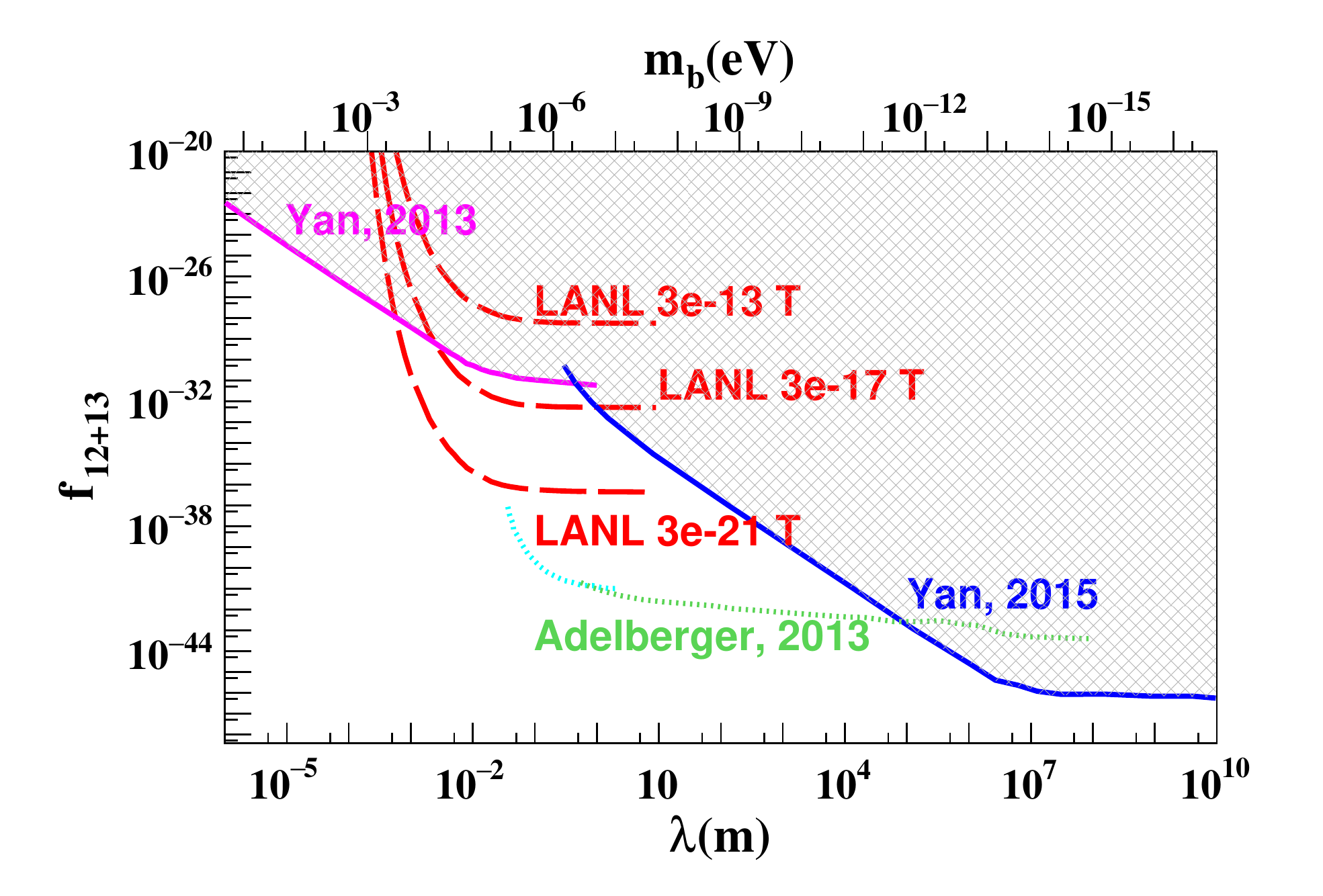}
\includegraphics[width=0.45\textwidth]{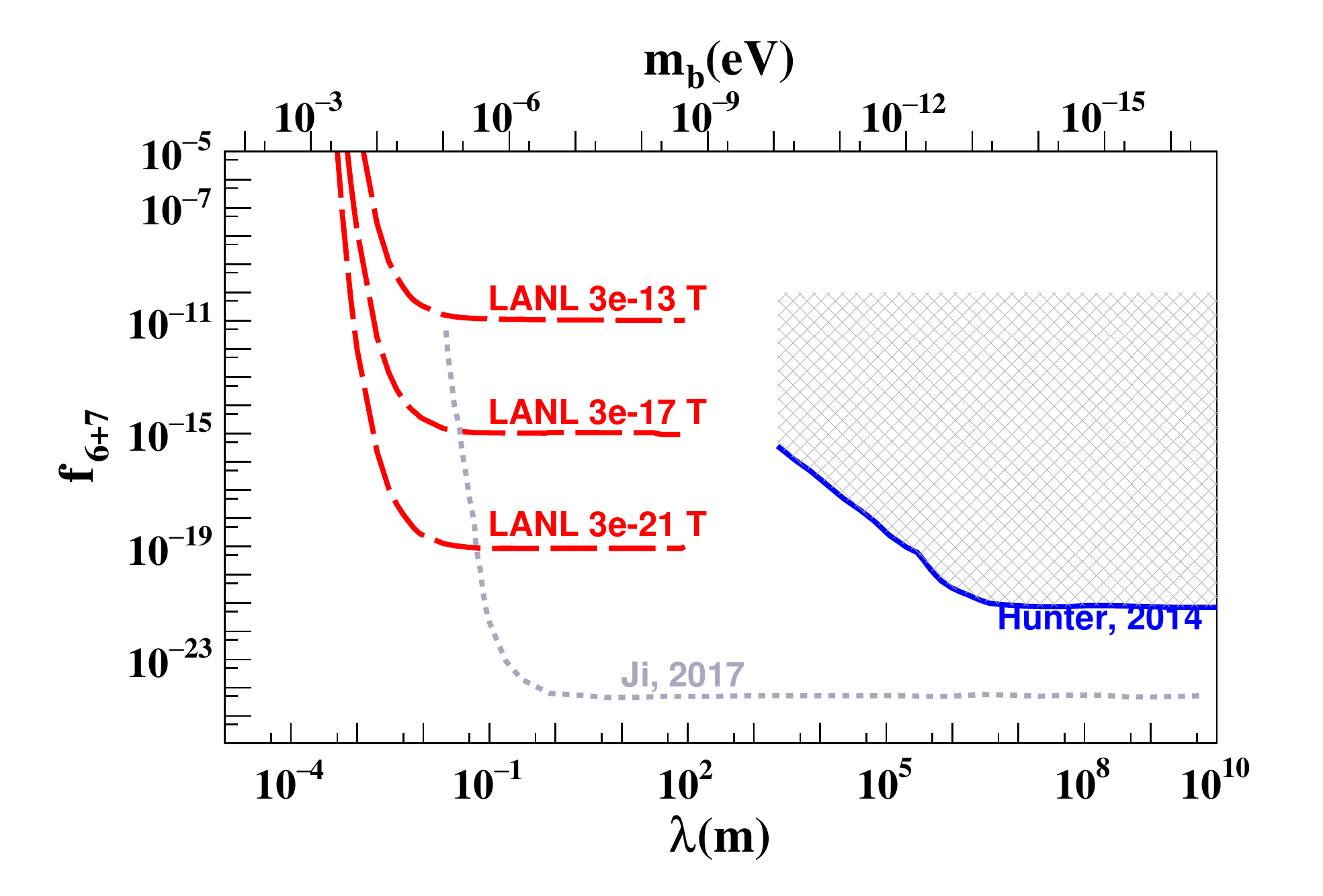}
\includegraphics[width=0.45\textwidth]{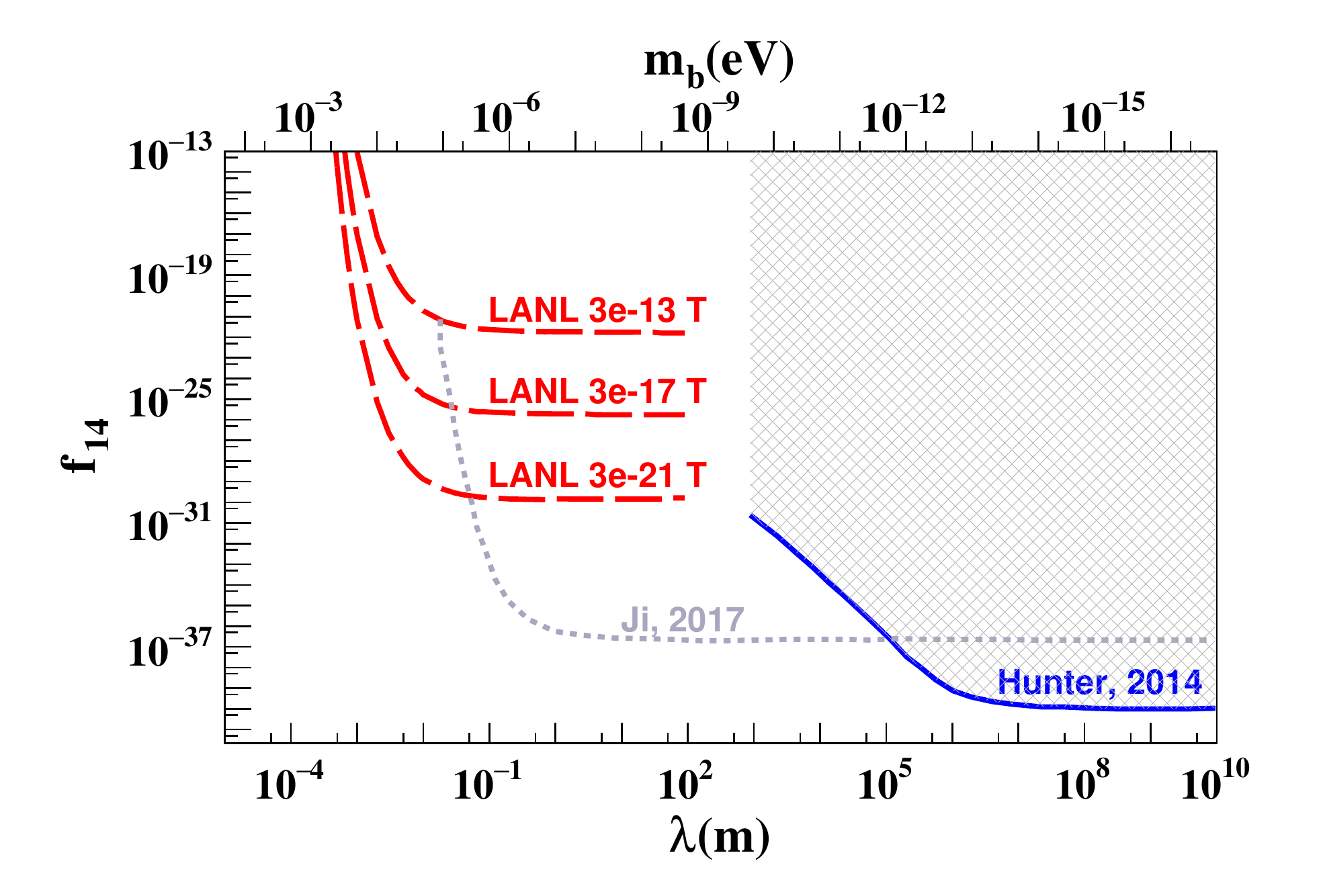}
\includegraphics[width=0.45\textwidth]{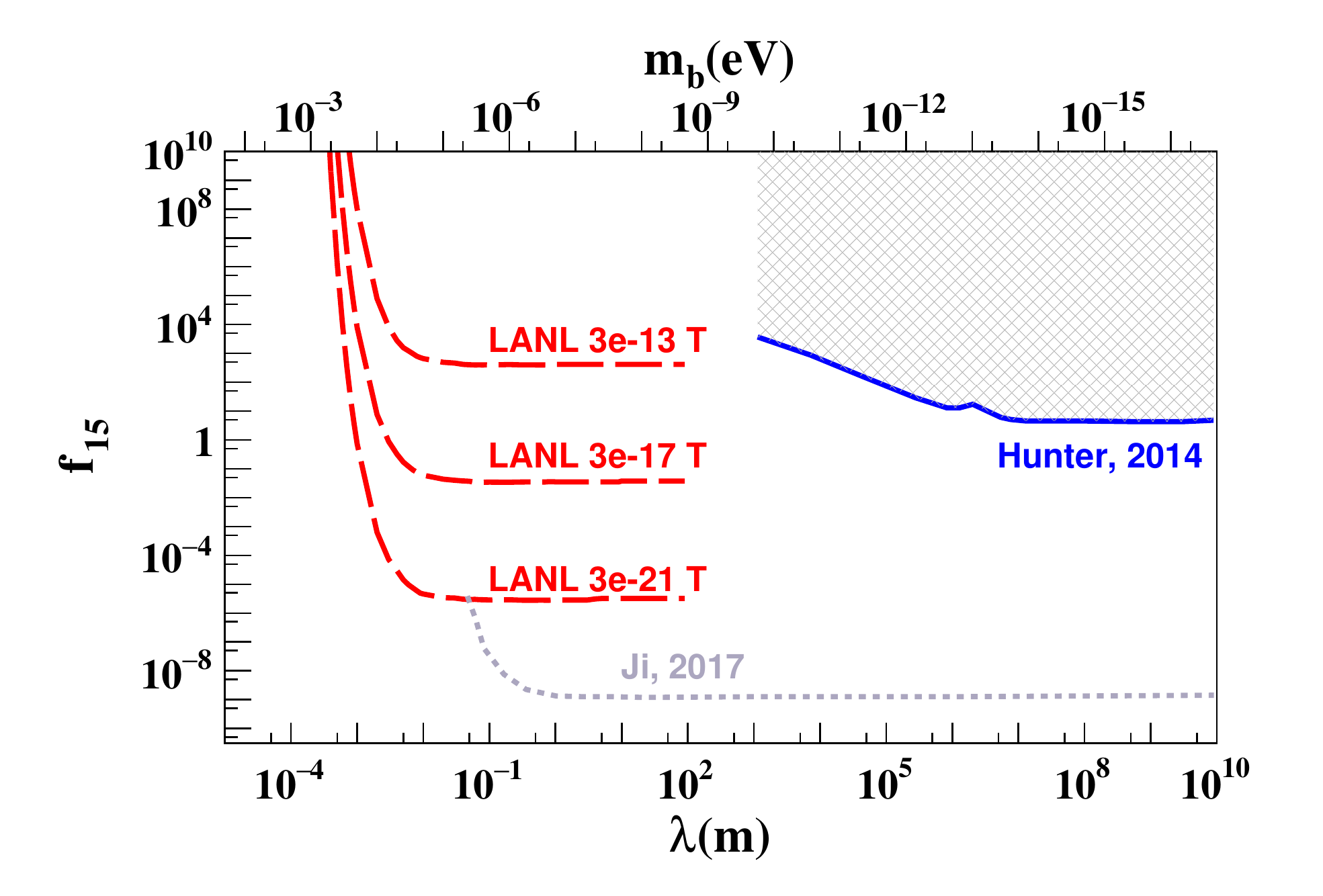}
\caption{The limits and the sensitivity estimation of $V_{4+5}$~\cite{Piegsa:2012}, $V_{12+13}$~\cite{Yan:2012wk,Yan:2015,Adelberger:2013}, $V_{6+7}$~\cite{Hunter:2014,Ji:2017}, $V_{14}$~\cite{Hunter:2014,Ji:2017}, and $V_{15}$~\cite{Hunter:2014,Ji:2017}.}
\label{fig:copuling_1}
\end{figure*}

\begin{figure*}[h]
\centering
\includegraphics[width=0.45\textwidth]{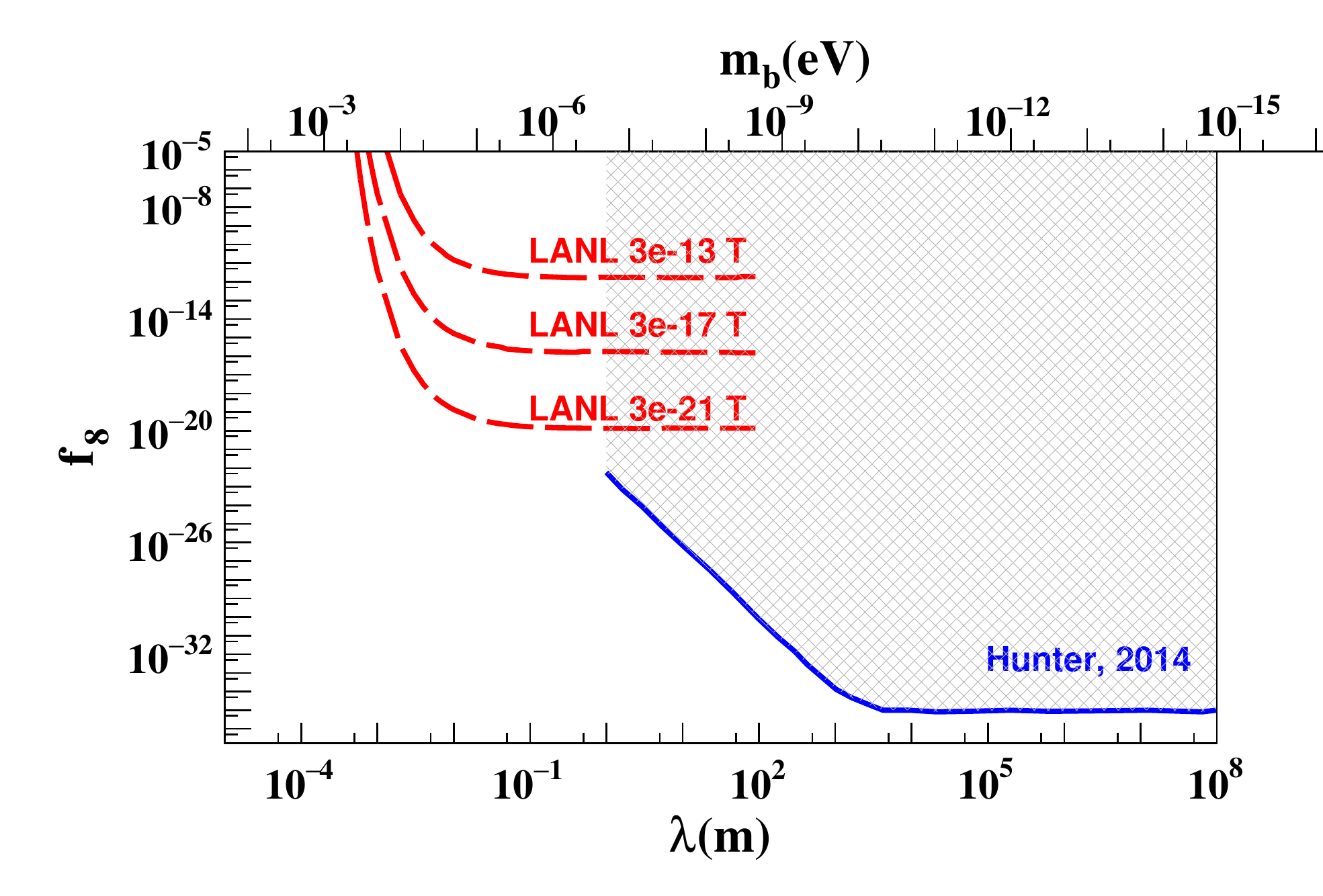}
\includegraphics[width=0.45\textwidth]{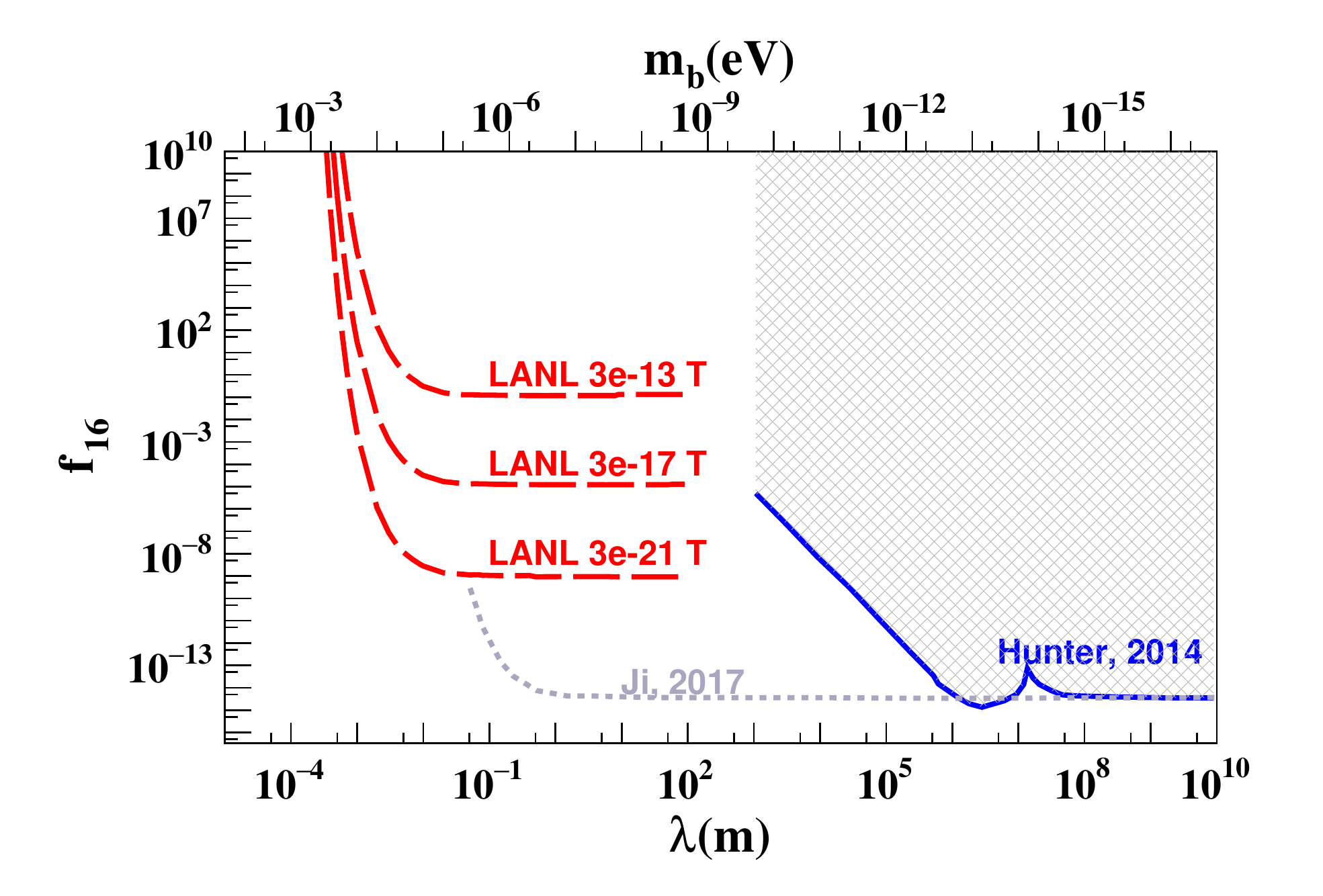}
\caption{The limits and the sensitivity estimation of  $V_{8}$~\cite{Hunter:2014}, and $V_{16}$~\cite{Hunter:2014,Ji:2017}.}
\label{fig:copuling_2}
\end{figure*}

The sensitivities of SDIs are estimated using the Monte Carlo method to average the interaction potentials in Eqs.~\ref{eq:v2}--~\ref{eq:v16} between $^3$He spins and particles in the mass ~\cite{Chu:2016, Kim:2019sry}. In Fig.~\ref{fig:copuling_0}--~\ref{fig:copuling_2}, we simply consider the $B_\text{eff}$ sensitivity of $3\times 10^{-13}$  T, $3\times 10^{-17}$  T, and $3\times 10^{-21}$ T for each interaction.  Most experiments worked on $V_{9+10}$~\cite{Wineland:1991,Venema:1992,Youdin:1996,Petukhov:2010,Chu:2012cf,Bulatowicz:2013,Tullney:2013,Guigue:2015fyt,Lee:2018vaq} and the resonance method at room temperature has a significant opportunity to improve the current constraints and even overcome the astrophysics constraints~\cite{Raffelt:2012sp}. In $V_{4+5}$, the only limit for the nuclear spin-dependent interaction was done using neutron beams with Ramsey’s technique of separated oscillating fields~\cite{Piegsa:2012,Haddock_2018}. We expect to improve the constraint with the frequency method or the resonance method. In $V_{12+13}$, the resonance method can also significantly improve the current constraints using neutron beams~\cite{Yan:2012wk}, the relaxation of polarized $^3$He spin relaxation~\cite{Yan:2015}, and the combination of different experiments~\cite{Adelberger:2013}. For spin-spin-dependent interactions, we only consider nucleon-electron interactions using polarized nuclear spins of $^3$He and polarized electron spins of DyIG. In $V_{3}$, the only constraint was done using $^{8}\text{Be}^{+}$ ion stored in Penning ion trap with polarized electron spins of a magnet~\cite{Wineland:1991} and the resonance method should improve the constraints. In $V_{2}, V_{11}, V_{6+7}, V_{8}, V_{14}, V_{15}, V_{16}$, the constraints between polarized nucleons and polarized electrons are rare. The only constraints were all done using the $^{199}$Hg-Cs comagnetometer with polarized electron spins of Earth~\cite{Hunter:2013, Hunter:2014} for possible long-range interactions. There is no constraint in the region of 1 m for the spin-spin-dependent interactions between nucleons and electrons. Using polarized $^{3}$He with the frequency method or the resonance method could be the first one to explore this interaction length range.

In conclusion, we estimate the sensitivity using a polarized $^{3}$He target with the frequency method and the resonance method~\cite{Arvanitaki:2014dfa} to search for the exotic spin-dependent interactions for polarized nucleons. Our calculations of the projected experimental sensitivity
showed that the experiments are sensitive to the interaction range of $10^{-2}$ to
$10^{-4}$ m. The resonance method especially has a significant potential to improve the current constraints. However, the reduction factor should be carefully calculated in the future for real experiments.

The authors thank P. E. Magnelind for his feedback. Research presented in this article was supported by the Laboratory Directed Research and Development program of Los Alamos National Laboratory under project number 20180129ER.

\bibliography{main}

\begin{thebibliography}{46}%
\makeatletter
\providecommand \@ifxundefined [1]{%
 \@ifx{#1\undefined}
}%
\providecommand \@ifnum [1]{%
 \ifnum #1\expandafter \@firstoftwo
 \else \expandafter \@secondoftwo
 \fi
}%
\providecommand \@ifx [1]{%
 \ifx #1\expandafter \@firstoftwo
 \else \expandafter \@secondoftwo
 \fi
}%
\providecommand \natexlab [1]{#1}%
\providecommand \enquote  [1]{``#1''}%
\providecommand \bibnamefont  [1]{#1}%
\providecommand \bibfnamefont [1]{#1}%
\providecommand \citenamefont [1]{#1}%
\providecommand \href@noop [0]{\@secondoftwo}%
\providecommand \href [0]{\begingroup \@sanitize@url \@href}%
\providecommand \@href[1]{\@@startlink{#1}\@@href}%
\providecommand \@@href[1]{\endgroup#1\@@endlink}%
\providecommand \@sanitize@url [0]{\catcode `\\12\catcode `\$12\catcode
  `\&12\catcode `\#12\catcode `\^12\catcode `\_12\catcode `\%12\relax}%
\providecommand \@@startlink[1]{}%
\providecommand \@@endlink[0]{}%
\providecommand \url  [0]{\begingroup\@sanitize@url \@url }%
\providecommand \@url [1]{\endgroup\@href {#1}{\urlprefix }}%
\providecommand \urlprefix  [0]{URL }%
\providecommand \Eprint [0]{\href }%
\providecommand \doibase [0]{http://dx.doi.org/}%
\providecommand \selectlanguage [0]{\@gobble}%
\providecommand \bibinfo  [0]{\@secondoftwo}%
\providecommand \bibfield  [0]{\@secondoftwo}%
\providecommand \translation [1]{[#1]}%
\providecommand \BibitemOpen [0]{}%
\providecommand \bibitemStop [0]{}%
\providecommand \bibitemNoStop [0]{.\EOS\space}%
\providecommand \EOS [0]{\spacefactor3000\relax}%
\providecommand \BibitemShut  [1]{\csname bibitem#1\endcsname}%
\let\auto@bib@innerbib\@empty
\bibitem [{\citenamefont {Commins}(2012)}]{Commins:2012}%
  \BibitemOpen
  \bibfield  {author} {\bibinfo {author} {\bibfnamefont {E.~D.}\ \bibnamefont
  {Commins}},\ }\href {\doibase 10.1146/annurev-nucl-102711-094908} {\bibfield
  {journal} {\bibinfo  {journal} {Annual Review of Nuclear and Particle
  Science}\ }\textbf {\bibinfo {volume} {62}},\ \bibinfo {pages} {133}
  (\bibinfo {year} {2012})}\BibitemShut {NoStop}%
\bibitem [{\citenamefont {Moody}\ and\ \citenamefont
  {Wilczek}(1984)}]{Moody:1984}%
  \BibitemOpen
  \bibfield  {author} {\bibinfo {author} {\bibfnamefont {J.~E.}\ \bibnamefont
  {Moody}}\ and\ \bibinfo {author} {\bibfnamefont {F.}~\bibnamefont
  {Wilczek}},\ }\href {\doibase 10.1103/PhysRevD.30.130} {\bibfield  {journal}
  {\bibinfo  {journal} {Phys. Rev. D}\ }\textbf {\bibinfo {volume} {30}},\
  \bibinfo {pages} {130} (\bibinfo {year} {1984})}\BibitemShut {NoStop}%
\bibitem [{\citenamefont {Dobrescu}\ and\ \citenamefont
  {Mocioiu}(2006)}]{Dobrescu:2006au}%
  \BibitemOpen
  \bibfield  {author} {\bibinfo {author} {\bibfnamefont {B.~A.}\ \bibnamefont
  {Dobrescu}}\ and\ \bibinfo {author} {\bibfnamefont {I.}~\bibnamefont
  {Mocioiu}},\ }\href {\doibase 10.1088/1126-6708/2006/11/005} {\bibfield
  {journal} {\bibinfo  {journal} {JHEP}\ }\textbf {\bibinfo {volume} {11}},\
  \bibinfo {pages} {005} (\bibinfo {year} {2006})}\BibitemShut {NoStop}%
\bibitem [{\citenamefont {Peccei}\ and\ \citenamefont
  {Quinn}(1977)}]{Peccei:1977}%
  \BibitemOpen
  \bibfield  {author} {\bibinfo {author} {\bibfnamefont {R.~D.}\ \bibnamefont
  {Peccei}}\ and\ \bibinfo {author} {\bibfnamefont {H.~R.}\ \bibnamefont
  {Quinn}},\ }\href {\doibase 10.1103/PhysRevLett.38.1440} {\bibfield
  {journal} {\bibinfo  {journal} {Phys. Rev. Lett.}\ }\textbf {\bibinfo
  {volume} {38}},\ \bibinfo {pages} {1440} (\bibinfo {year}
  {1977})}\BibitemShut {NoStop}%
\bibitem [{\citenamefont {Duffy}\ and\ \citenamefont {van
  Bibber}(2009)}]{Duffy:2009ig}%
  \BibitemOpen
  \bibfield  {author} {\bibinfo {author} {\bibfnamefont {L.~D.}\ \bibnamefont
  {Duffy}}\ and\ \bibinfo {author} {\bibfnamefont {K.}~\bibnamefont {van
  Bibber}},\ }\href {\doibase 10.1088/1367-2630/11/10/105008} {\bibfield
  {journal} {\bibinfo  {journal} {New J. Phys.}\ }\textbf {\bibinfo {volume}
  {11}},\ \bibinfo {pages} {105008} (\bibinfo {year} {2009})}\BibitemShut
  {NoStop}%
\bibitem [{\citenamefont {Arvanitaki}\ \emph {et~al.}(2010)\citenamefont
  {Arvanitaki}, \citenamefont {Dimopoulos}, \citenamefont {Dubovsky},
  \citenamefont {Kaloper},\ and\ \citenamefont
  {March-Russell}}]{Arvanitaki:2010}%
  \BibitemOpen
  \bibfield  {author} {\bibinfo {author} {\bibfnamefont {A.}~\bibnamefont
  {Arvanitaki}}, \bibinfo {author} {\bibfnamefont {S.}~\bibnamefont
  {Dimopoulos}}, \bibinfo {author} {\bibfnamefont {S.}~\bibnamefont
  {Dubovsky}}, \bibinfo {author} {\bibfnamefont {N.}~\bibnamefont {Kaloper}}, \
  and\ \bibinfo {author} {\bibfnamefont {J.}~\bibnamefont {March-Russell}},\
  }\href {\doibase 10.1103/PhysRevD.81.123530} {\bibfield  {journal} {\bibinfo
  {journal} {Phys. Rev. D}\ }\textbf {\bibinfo {volume} {81}},\ \bibinfo
  {pages} {123530} (\bibinfo {year} {2010})}\BibitemShut {NoStop}%
\bibitem [{\citenamefont {Graham}\ \emph {et~al.}(2015)\citenamefont {Graham},
  \citenamefont {Kaplan},\ and\ \citenamefont {Rajendran}}]{Graham:2015}%
  \BibitemOpen
  \bibfield  {author} {\bibinfo {author} {\bibfnamefont {P.~W.}\ \bibnamefont
  {Graham}}, \bibinfo {author} {\bibfnamefont {D.~E.}\ \bibnamefont {Kaplan}},
  \ and\ \bibinfo {author} {\bibfnamefont {S.}~\bibnamefont {Rajendran}},\
  }\href {\doibase 10.1103/PhysRevLett.115.221801} {\bibfield  {journal}
  {\bibinfo  {journal} {Phys. Rev. Lett.}\ }\textbf {\bibinfo {volume} {115}},\
  \bibinfo {pages} {221801} (\bibinfo {year} {2015})}\BibitemShut {NoStop}%
\bibitem [{\citenamefont {Flambaum}\ \emph {et~al.}(2009)\citenamefont
  {Flambaum}, \citenamefont {Lambert},\ and\ \citenamefont
  {Pospelov}}]{Flambaum:2009}%
  \BibitemOpen
  \bibfield  {author} {\bibinfo {author} {\bibfnamefont {V.}~\bibnamefont
  {Flambaum}}, \bibinfo {author} {\bibfnamefont {S.}~\bibnamefont {Lambert}}, \
  and\ \bibinfo {author} {\bibfnamefont {M.}~\bibnamefont {Pospelov}},\ }\href
  {\doibase 10.1103/PhysRevD.80.105021} {\bibfield  {journal} {\bibinfo
  {journal} {Phys. Rev. D}\ }\textbf {\bibinfo {volume} {80}},\ \bibinfo
  {pages} {105021} (\bibinfo {year} {2009})}\BibitemShut {NoStop}%
\bibitem [{\citenamefont {Georgi}(2007)}]{Georgi:2007}%
  \BibitemOpen
  \bibfield  {author} {\bibinfo {author} {\bibfnamefont {H.}~\bibnamefont
  {Georgi}},\ }\href {\doibase 10.1103/PhysRevLett.98.221601} {\bibfield
  {journal} {\bibinfo  {journal} {Phys. Rev. Lett.}\ }\textbf {\bibinfo
  {volume} {98}},\ \bibinfo {pages} {221601} (\bibinfo {year}
  {2007})}\BibitemShut {NoStop}%
\bibitem [{\citenamefont {Appelquist}\ \emph {et~al.}(2003)\citenamefont
  {Appelquist}, \citenamefont {Dobrescu},\ and\ \citenamefont
  {Hopper}}]{Appelquist:2003}%
  \BibitemOpen
  \bibfield  {author} {\bibinfo {author} {\bibfnamefont {T.}~\bibnamefont
  {Appelquist}}, \bibinfo {author} {\bibfnamefont {B.~A.}\ \bibnamefont
  {Dobrescu}}, \ and\ \bibinfo {author} {\bibfnamefont {A.~R.}\ \bibnamefont
  {Hopper}},\ }\href {\doibase 10.1103/PhysRevD.68.035012} {\bibfield
  {journal} {\bibinfo  {journal} {Phys. Rev. D}\ }\textbf {\bibinfo {volume}
  {68}},\ \bibinfo {pages} {035012} (\bibinfo {year} {2003})}\BibitemShut
  {NoStop}%
\bibitem [{\citenamefont {Dobrescu}(2005)}]{Dobrescu:2005}%
  \BibitemOpen
  \bibfield  {author} {\bibinfo {author} {\bibfnamefont {B.~A.}\ \bibnamefont
  {Dobrescu}},\ }\href {\doibase 10.1103/PhysRevLett.94.151802} {\bibfield
  {journal} {\bibinfo  {journal} {Phys. Rev. Lett.}\ }\textbf {\bibinfo
  {volume} {94}},\ \bibinfo {pages} {151802} (\bibinfo {year}
  {2005})}\BibitemShut {NoStop}%
\bibitem [{\citenamefont {Ackerman}\ \emph {et~al.}(2009)\citenamefont
  {Ackerman}, \citenamefont {Buckley}, \citenamefont {Carroll},\ and\
  \citenamefont {Kamionkowski}}]{Ackerman:2009}%
  \BibitemOpen
  \bibfield  {author} {\bibinfo {author} {\bibfnamefont {L.}~\bibnamefont
  {Ackerman}}, \bibinfo {author} {\bibfnamefont {M.~R.}\ \bibnamefont
  {Buckley}}, \bibinfo {author} {\bibfnamefont {S.~M.}\ \bibnamefont
  {Carroll}}, \ and\ \bibinfo {author} {\bibfnamefont {M.}~\bibnamefont
  {Kamionkowski}},\ }\href {\doibase 10.1103/PhysRevD.79.023519} {\bibfield
  {journal} {\bibinfo  {journal} {Phys. Rev. D}\ }\textbf {\bibinfo {volume}
  {79}},\ \bibinfo {pages} {023519} (\bibinfo {year} {2009})}\BibitemShut
  {NoStop}%
\bibitem [{\citenamefont {Safronova}\ \emph {et~al.}(2018)\citenamefont
  {Safronova}, \citenamefont {Budker}, \citenamefont {DeMille}, \citenamefont
  {Kimball}, \citenamefont {Derevianko},\ and\ \citenamefont
  {Clark}}]{Safronova:2017xyt}%
  \BibitemOpen
  \bibfield  {author} {\bibinfo {author} {\bibfnamefont {M.~S.}\ \bibnamefont
  {Safronova}}, \bibinfo {author} {\bibfnamefont {D.}~\bibnamefont {Budker}},
  \bibinfo {author} {\bibfnamefont {D.}~\bibnamefont {DeMille}}, \bibinfo
  {author} {\bibfnamefont {D.~F.~J.}\ \bibnamefont {Kimball}}, \bibinfo
  {author} {\bibfnamefont {A.}~\bibnamefont {Derevianko}}, \ and\ \bibinfo
  {author} {\bibfnamefont {C.~W.}\ \bibnamefont {Clark}},\ }\href {\doibase
  10.1103/RevModPhys.90.025008} {\bibfield  {journal} {\bibinfo  {journal}
  {Rev. Mod. Phys.}\ }\textbf {\bibinfo {volume} {90}},\ \bibinfo {pages}
  {025008} (\bibinfo {year} {2018})}\BibitemShut {NoStop}%
\bibitem [{\citenamefont {Leslie}\ \emph {et~al.}(2014)\citenamefont {Leslie},
  \citenamefont {Weisman}, \citenamefont {Khatiwada},\ and\ \citenamefont
  {Long}}]{Leslie:2014mua}%
  \BibitemOpen
  \bibfield  {author} {\bibinfo {author} {\bibfnamefont {T.~M.}\ \bibnamefont
  {Leslie}}, \bibinfo {author} {\bibfnamefont {E.}~\bibnamefont {Weisman}},
  \bibinfo {author} {\bibfnamefont {R.}~\bibnamefont {Khatiwada}}, \ and\
  \bibinfo {author} {\bibfnamefont {J.~C.}\ \bibnamefont {Long}},\ }\href
  {\doibase 10.1103/PhysRevD.89.114022} {\bibfield  {journal} {\bibinfo
  {journal} {Phys. Rev.}\ }\textbf {\bibinfo {volume} {D89}},\ \bibinfo {pages}
  {114022} (\bibinfo {year} {2014})}\BibitemShut {NoStop}%
\bibitem [{\citenamefont {Fadeev}\ \emph
  {et~al.}(2019{\natexlab{a}})\citenamefont {Fadeev}, \citenamefont {Stadnik},
  \citenamefont {Ficek}, \citenamefont {Kozlov}, \citenamefont {Flambaum},\
  and\ \citenamefont {Budker}}]{Fadeev:2018rfl}%
  \BibitemOpen
  \bibfield  {author} {\bibinfo {author} {\bibfnamefont {P.}~\bibnamefont
  {Fadeev}}, \bibinfo {author} {\bibfnamefont {Y.~V.}\ \bibnamefont {Stadnik}},
  \bibinfo {author} {\bibfnamefont {F.}~\bibnamefont {Ficek}}, \bibinfo
  {author} {\bibfnamefont {M.~G.}\ \bibnamefont {Kozlov}}, \bibinfo {author}
  {\bibfnamefont {V.~V.}\ \bibnamefont {Flambaum}}, \ and\ \bibinfo {author}
  {\bibfnamefont {D.}~\bibnamefont {Budker}},\ }\href {\doibase
  10.1103/PhysRevA.99.022113} {\bibfield  {journal} {\bibinfo  {journal} {Phys.
  Rev.}\ }\textbf {\bibinfo {volume} {A99}},\ \bibinfo {pages} {022113}
  (\bibinfo {year} {2019}{\natexlab{a}})}\BibitemShut {NoStop}%
\bibitem [{\citenamefont {Fadeev}\ \emph
  {et~al.}(2019{\natexlab{b}})\citenamefont {Fadeev}, \citenamefont {Ficek},
  \citenamefont {Kozlov}, \citenamefont {Budker},\ and\ \citenamefont
  {Flambaum}}]{Fadeev:2019jzi}%
  \BibitemOpen
  \bibfield  {author} {\bibinfo {author} {\bibfnamefont {P.}~\bibnamefont
  {Fadeev}}, \bibinfo {author} {\bibfnamefont {F.}~\bibnamefont {Ficek}},
  \bibinfo {author} {\bibfnamefont {M.~G.}\ \bibnamefont {Kozlov}}, \bibinfo
  {author} {\bibfnamefont {D.}~\bibnamefont {Budker}}, \ and\ \bibinfo {author}
  {\bibfnamefont {V.~V.}\ \bibnamefont {Flambaum}},\ }\href@noop {} {\
  (\bibinfo {year} {2019}{\natexlab{b}})},\ \Eprint
  {http://arxiv.org/abs/1911.05816} {arXiv:1911.05816 [hep-ph]} \BibitemShut
  {NoStop}%
\bibitem [{\citenamefont {Chu}\ \emph {et~al.}(2016)\citenamefont {Chu},
  \citenamefont {Kim},\ and\ \citenamefont {Savukov}}]{Chu:2016}%
  \BibitemOpen
  \bibfield  {author} {\bibinfo {author} {\bibfnamefont {P.-H.}\ \bibnamefont
  {Chu}}, \bibinfo {author} {\bibfnamefont {Y.~J.}\ \bibnamefont {Kim}}, \ and\
  \bibinfo {author} {\bibfnamefont {I.}~\bibnamefont {Savukov}},\ }\href
  {\doibase 10.1103/PhysRevD.94.036002} {\bibfield  {journal} {\bibinfo
  {journal} {Phys. Rev. D}\ }\textbf {\bibinfo {volume} {94}},\ \bibinfo
  {pages} {036002} (\bibinfo {year} {2016})}\BibitemShut {NoStop}%
\bibitem [{\citenamefont {Yan}\ and\ \citenamefont {Snow}(2013)}]{Yan:2012wk}%
  \BibitemOpen
  \bibfield  {author} {\bibinfo {author} {\bibfnamefont {H.}~\bibnamefont
  {Yan}}\ and\ \bibinfo {author} {\bibfnamefont {W.~M.}\ \bibnamefont {Snow}},\
  }\href {\doibase 10.1103/PhysRevLett.110.082003} {\bibfield  {journal}
  {\bibinfo  {journal} {Phys. Rev. Lett.}\ }\textbf {\bibinfo {volume} {110}},\
  \bibinfo {pages} {082003} (\bibinfo {year} {2013})},\ \Eprint
  {http://arxiv.org/abs/1211.6523} {arXiv:1211.6523 [nucl-ex]} \BibitemShut
  {NoStop}%
\bibitem [{\citenamefont {Yan}\ \emph {et~al.}(2015)\citenamefont {Yan},
  \citenamefont {Sun}, \citenamefont {Peng}, \citenamefont {Zhang},
  \citenamefont {Fu}, \citenamefont {Guo},\ and\ \citenamefont
  {Liu}}]{Yan:2015}%
  \BibitemOpen
  \bibfield  {author} {\bibinfo {author} {\bibfnamefont {H.}~\bibnamefont
  {Yan}}, \bibinfo {author} {\bibfnamefont {G.~A.}\ \bibnamefont {Sun}},
  \bibinfo {author} {\bibfnamefont {S.~M.}\ \bibnamefont {Peng}}, \bibinfo
  {author} {\bibfnamefont {Y.}~\bibnamefont {Zhang}}, \bibinfo {author}
  {\bibfnamefont {C.}~\bibnamefont {Fu}}, \bibinfo {author} {\bibfnamefont
  {H.}~\bibnamefont {Guo}}, \ and\ \bibinfo {author} {\bibfnamefont {B.~Q.}\
  \bibnamefont {Liu}},\ }\href {\doibase 10.1103/PhysRevLett.115.182001}
  {\bibfield  {journal} {\bibinfo  {journal} {Phys. Rev. Lett.}\ }\textbf
  {\bibinfo {volume} {115}},\ \bibinfo {pages} {182001} (\bibinfo {year}
  {2015})}\BibitemShut {NoStop}%
\bibitem [{\citenamefont {Adelberger}\ and\ \citenamefont
  {Wagner}(2013)}]{Adelberger:2013}%
  \BibitemOpen
  \bibfield  {author} {\bibinfo {author} {\bibfnamefont {E.~G.}\ \bibnamefont
  {Adelberger}}\ and\ \bibinfo {author} {\bibfnamefont {T.~A.}\ \bibnamefont
  {Wagner}},\ }\href {\doibase 10.1103/PhysRevD.88.031101} {\bibfield
  {journal} {\bibinfo  {journal} {Phys. Rev. D}\ }\textbf {\bibinfo {volume}
  {88}},\ \bibinfo {pages} {031101} (\bibinfo {year} {2013})}\BibitemShut
  {NoStop}%
\bibitem [{\citenamefont {Piegsa}\ and\ \citenamefont
  {Pignol}(2012)}]{Piegsa:2012}%
  \BibitemOpen
  \bibfield  {author} {\bibinfo {author} {\bibfnamefont {F.~M.}\ \bibnamefont
  {Piegsa}}\ and\ \bibinfo {author} {\bibfnamefont {G.}~\bibnamefont
  {Pignol}},\ }\href {\doibase 10.1103/PhysRevLett.108.181801} {\bibfield
  {journal} {\bibinfo  {journal} {Phys. Rev. Lett.}\ }\textbf {\bibinfo
  {volume} {108}},\ \bibinfo {pages} {181801} (\bibinfo {year}
  {2012})}\BibitemShut {NoStop}%
\bibitem [{\citenamefont {Haddock}\ \emph {et~al.}(2018)\citenamefont
  {Haddock}, \citenamefont {Amadio}, \citenamefont {Anderson}, \citenamefont
  {Barrón-Palos}, \citenamefont {Crawford}, \citenamefont {Crawford},
  \citenamefont {Esposito}, \citenamefont {Fox}, \citenamefont {Francis},
  \citenamefont {Fry},\ and\ \citenamefont {et~al.}}]{Haddock_2018}%
  \BibitemOpen
  \bibfield  {author} {\bibinfo {author} {\bibfnamefont {C.}~\bibnamefont
  {Haddock}}, \bibinfo {author} {\bibfnamefont {J.}~\bibnamefont {Amadio}},
  \bibinfo {author} {\bibfnamefont {E.}~\bibnamefont {Anderson}}, \bibinfo
  {author} {\bibfnamefont {L.}~\bibnamefont {Barrón-Palos}}, \bibinfo {author}
  {\bibfnamefont {B.}~\bibnamefont {Crawford}}, \bibinfo {author}
  {\bibfnamefont {C.}~\bibnamefont {Crawford}}, \bibinfo {author}
  {\bibfnamefont {D.}~\bibnamefont {Esposito}}, \bibinfo {author}
  {\bibfnamefont {W.}~\bibnamefont {Fox}}, \bibinfo {author} {\bibfnamefont
  {I.}~\bibnamefont {Francis}}, \bibinfo {author} {\bibfnamefont
  {J.}~\bibnamefont {Fry}}, \ and\ \bibinfo {author} {\bibnamefont {et~al.}},\
  }\href {\doibase 10.1016/j.physletb.2018.06.066} {\bibfield  {journal}
  {\bibinfo  {journal} {Physics Letters B}\ }\textbf {\bibinfo {volume}
  {783}},\ \bibinfo {pages} {227–233} (\bibinfo {year} {2018})}\BibitemShut
  {NoStop}%
\bibitem [{\citenamefont {Kim}\ \emph {et~al.}(2018)\citenamefont {Kim},
  \citenamefont {Chu},\ and\ \citenamefont {Savukov}}]{Kim:2017yen}%
  \BibitemOpen
  \bibfield  {author} {\bibinfo {author} {\bibfnamefont {Y.~J.}\ \bibnamefont
  {Kim}}, \bibinfo {author} {\bibfnamefont {P.-H.}\ \bibnamefont {Chu}}, \ and\
  \bibinfo {author} {\bibfnamefont {I.}~\bibnamefont {Savukov}},\ }\href
  {\doibase 10.1103/PhysRevLett.121.091802} {\bibfield  {journal} {\bibinfo
  {journal} {Phys. Rev. Lett.}\ }\textbf {\bibinfo {volume} {121}},\ \bibinfo
  {pages} {091802} (\bibinfo {year} {2018})},\ \Eprint
  {http://arxiv.org/abs/1702.02974} {arXiv:1702.02974 [physics.ins-det]}
  \BibitemShut {NoStop}%
\bibitem [{\citenamefont {Kim}\ \emph {et~al.}(2019)\citenamefont {Kim},
  \citenamefont {Chu}, \citenamefont {Savukov},\ and\ \citenamefont
  {Newman}}]{Kim:2019sry}%
  \BibitemOpen
  \bibfield  {author} {\bibinfo {author} {\bibfnamefont {Y.~J.}\ \bibnamefont
  {Kim}}, \bibinfo {author} {\bibfnamefont {P.-H.}\ \bibnamefont {Chu}},
  \bibinfo {author} {\bibfnamefont {I.}~\bibnamefont {Savukov}}, \ and\
  \bibinfo {author} {\bibfnamefont {S.}~\bibnamefont {Newman}},\ }\href
  {\doibase 10.1038/s41467-019-10169-1} {\bibfield  {journal} {\bibinfo
  {journal} {Nature Commun.}\ }\textbf {\bibinfo {volume} {10}},\ \bibinfo
  {pages} {2245} (\bibinfo {year} {2019})},\ \Eprint
  {http://arxiv.org/abs/1902.00128} {arXiv:1902.00128 [hep-ex]} \BibitemShut
  {NoStop}%
\bibitem [{\citenamefont {Hunter}\ \emph {et~al.}(2013)\citenamefont {Hunter},
  \citenamefont {Gordon}, \citenamefont {Peck}, \citenamefont {Ang},\ and\
  \citenamefont {Lin}}]{Hunter:2013}%
  \BibitemOpen
  \bibfield  {author} {\bibinfo {author} {\bibfnamefont {L.}~\bibnamefont
  {Hunter}}, \bibinfo {author} {\bibfnamefont {J.}~\bibnamefont {Gordon}},
  \bibinfo {author} {\bibfnamefont {S.}~\bibnamefont {Peck}}, \bibinfo {author}
  {\bibfnamefont {D.}~\bibnamefont {Ang}}, \ and\ \bibinfo {author}
  {\bibfnamefont {J.-F.}\ \bibnamefont {Lin}},\ }\href {\doibase
  10.1126/science.1227460} {\bibfield  {journal} {\bibinfo  {journal}
  {Science}\ }\textbf {\bibinfo {volume} {339}},\ \bibinfo {pages} {928}
  (\bibinfo {year} {2013})}\BibitemShut {NoStop}%
\bibitem [{\citenamefont {Hunter}\ and\ \citenamefont
  {Ang}(2014)}]{Hunter:2014}%
  \BibitemOpen
  \bibfield  {author} {\bibinfo {author} {\bibfnamefont {L.~R.}\ \bibnamefont
  {Hunter}}\ and\ \bibinfo {author} {\bibfnamefont {D.~G.}\ \bibnamefont
  {Ang}},\ }\href {\doibase 10.1103/PhysRevLett.112.091803} {\bibfield
  {journal} {\bibinfo  {journal} {Phys. Rev. Lett.}\ }\textbf {\bibinfo
  {volume} {112}},\ \bibinfo {pages} {091803} (\bibinfo {year}
  {2014})}\BibitemShut {NoStop}%
\bibitem [{\citenamefont {Ji}\ \emph {et~al.}(2017)\citenamefont {Ji},
  \citenamefont {Fu},\ and\ \citenamefont {Gao}}]{Ji:2017}%
  \BibitemOpen
  \bibfield  {author} {\bibinfo {author} {\bibfnamefont {W.}~\bibnamefont
  {Ji}}, \bibinfo {author} {\bibfnamefont {C.~B.}\ \bibnamefont {Fu}}, \ and\
  \bibinfo {author} {\bibfnamefont {H.}~\bibnamefont {Gao}},\ }\href {\doibase
  10.1103/PhysRevD.95.075014} {\bibfield  {journal} {\bibinfo  {journal} {Phys.
  Rev. D}\ }\textbf {\bibinfo {volume} {95}},\ \bibinfo {pages} {075014}
  (\bibinfo {year} {2017})}\BibitemShut {NoStop}%
\bibitem [{\citenamefont {Ji}\ \emph {et~al.}(2018)\citenamefont {Ji},
  \citenamefont {Chen}, \citenamefont {Fu}, \citenamefont {Ding}, \citenamefont
  {Fang}, \citenamefont {Xiao}, \citenamefont {Wei},\ and\ \citenamefont
  {Yan}}]{Ji:2018}%
  \BibitemOpen
  \bibfield  {author} {\bibinfo {author} {\bibfnamefont {W.}~\bibnamefont
  {Ji}}, \bibinfo {author} {\bibfnamefont {Y.}~\bibnamefont {Chen}}, \bibinfo
  {author} {\bibfnamefont {C.}~\bibnamefont {Fu}}, \bibinfo {author}
  {\bibfnamefont {M.}~\bibnamefont {Ding}}, \bibinfo {author} {\bibfnamefont
  {J.}~\bibnamefont {Fang}}, \bibinfo {author} {\bibfnamefont {Z.}~\bibnamefont
  {Xiao}}, \bibinfo {author} {\bibfnamefont {K.}~\bibnamefont {Wei}}, \ and\
  \bibinfo {author} {\bibfnamefont {H.}~\bibnamefont {Yan}},\ }\href {\doibase
  10.1103/PhysRevLett.121.261803} {\bibfield  {journal} {\bibinfo  {journal}
  {Phys. Rev. Lett.}\ }\textbf {\bibinfo {volume} {121}},\ \bibinfo {pages}
  {261803} (\bibinfo {year} {2018})}\BibitemShut {NoStop}%
\bibitem [{\citenamefont {Chu}\ \emph {et~al.}(2015)\citenamefont {Chu},
  \citenamefont {Weisman}, \citenamefont {Liu},\ and\ \citenamefont
  {Long}}]{Chu:2015tha}%
  \BibitemOpen
  \bibfield  {author} {\bibinfo {author} {\bibfnamefont {P.-H.}\ \bibnamefont
  {Chu}}, \bibinfo {author} {\bibfnamefont {E.}~\bibnamefont {Weisman}},
  \bibinfo {author} {\bibfnamefont {C.-Y.}\ \bibnamefont {Liu}}, \ and\
  \bibinfo {author} {\bibfnamefont {J.~C.}\ \bibnamefont {Long}},\ }\href
  {\doibase 10.1103/PhysRevD.91.102006} {\bibfield  {journal} {\bibinfo
  {journal} {Phys. Rev.}\ }\textbf {\bibinfo {volume} {D91}},\ \bibinfo {pages}
  {102006} (\bibinfo {year} {2015})},\ \Eprint
  {http://arxiv.org/abs/1504.00552} {arXiv:1504.00552 [hep-ph]} \BibitemShut
  {NoStop}%
\bibitem [{\citenamefont {Chu}\ \emph {et~al.}(2013)\citenamefont {Chu} \emph
  {et~al.}}]{Chu:2012cf}%
  \BibitemOpen
  \bibfield  {author} {\bibinfo {author} {\bibfnamefont {P.~H.}\ \bibnamefont
  {Chu}} \emph {et~al.},\ }\href {\doibase 10.1103/PhysRevD.87.011105}
  {\bibfield  {journal} {\bibinfo  {journal} {Phys. Rev.}\ }\textbf {\bibinfo
  {volume} {D87}},\ \bibinfo {pages} {011105} (\bibinfo {year}
  {2013})}\BibitemShut {NoStop}%
\bibitem [{\citenamefont {Arvanitaki}\ and\ \citenamefont
  {Geraci}(2014)}]{Arvanitaki:2014dfa}%
  \BibitemOpen
  \bibfield  {author} {\bibinfo {author} {\bibfnamefont {A.}~\bibnamefont
  {Arvanitaki}}\ and\ \bibinfo {author} {\bibfnamefont {A.~A.}\ \bibnamefont
  {Geraci}},\ }\href {\doibase 10.1103/PhysRevLett.113.161801} {\bibfield
  {journal} {\bibinfo  {journal} {Phys. Rev. Lett.}\ }\textbf {\bibinfo
  {volume} {113}},\ \bibinfo {pages} {161801} (\bibinfo {year} {2014})},\
  \Eprint {http://arxiv.org/abs/1403.1290} {arXiv:1403.1290 [hep-ph]}
  \BibitemShut {NoStop}%
\bibitem [{\citenamefont {Gentile}\ \emph {et~al.}(2017)\citenamefont
  {Gentile}, \citenamefont {Nacher}, \citenamefont {Saam},\ and\ \citenamefont
  {Walker}}]{Gentile_2017}%
  \BibitemOpen
  \bibfield  {author} {\bibinfo {author} {\bibfnamefont {T.}~\bibnamefont
  {Gentile}}, \bibinfo {author} {\bibfnamefont {P.}~\bibnamefont {Nacher}},
  \bibinfo {author} {\bibfnamefont {B.}~\bibnamefont {Saam}}, \ and\ \bibinfo
  {author} {\bibfnamefont {T.}~\bibnamefont {Walker}},\ }\href {\doibase
  10.1103/revmodphys.89.045004} {\bibfield  {journal} {\bibinfo  {journal}
  {Reviews of Modern Physics}\ }\textbf {\bibinfo {volume} {89}} (\bibinfo
  {year} {2017}),\ 10.1103/revmodphys.89.045004}\BibitemShut {NoStop}%
\bibitem [{\citenamefont {Budker}\ and\ \citenamefont
  {Romalis}(2007)}]{Budker_2007}%
  \BibitemOpen
  \bibfield  {author} {\bibinfo {author} {\bibfnamefont {D.}~\bibnamefont
  {Budker}}\ and\ \bibinfo {author} {\bibfnamefont {M.}~\bibnamefont
  {Romalis}},\ }\href {\doibase 10.1038/nphys566} {\bibfield  {journal}
  {\bibinfo  {journal} {Nature Physics}\ }\textbf {\bibinfo {volume} {3}},\
  \bibinfo {pages} {227–234} (\bibinfo {year} {2007})}\BibitemShut {NoStop}%
\bibitem [{\citenamefont {Rosenberry}\ and\ \citenamefont
  {Chupp}(2001)}]{Rosenberry:2001}%
  \BibitemOpen
  \bibfield  {author} {\bibinfo {author} {\bibfnamefont {M.~A.}\ \bibnamefont
  {Rosenberry}}\ and\ \bibinfo {author} {\bibfnamefont {T.~E.}\ \bibnamefont
  {Chupp}},\ }\href {\doibase 10.1103/PhysRevLett.86.22} {\bibfield  {journal}
  {\bibinfo  {journal} {Phys. Rev. Lett.}\ }\textbf {\bibinfo {volume} {86}},\
  \bibinfo {pages} {22} (\bibinfo {year} {2001})}\BibitemShut {NoStop}%
\bibitem [{\citenamefont {Tullney}\ \emph {et~al.}(2013)\citenamefont
  {Tullney}, \citenamefont {Allmendinger}, \citenamefont {Burghoff},
  \citenamefont {Heil}, \citenamefont {Karpuk}, \citenamefont {Kilian},
  \citenamefont {Knappe-Gr\"uneberg}, \citenamefont {M\"uller}, \citenamefont
  {Schmidt}, \citenamefont {Schnabel}, \citenamefont {Seifert}, \citenamefont
  {Sobolev},\ and\ \citenamefont {Trahms}}]{Tullney:2013}%
  \BibitemOpen
  \bibfield  {author} {\bibinfo {author} {\bibfnamefont {K.}~\bibnamefont
  {Tullney}}, \bibinfo {author} {\bibfnamefont {F.}~\bibnamefont
  {Allmendinger}}, \bibinfo {author} {\bibfnamefont {M.}~\bibnamefont
  {Burghoff}}, \bibinfo {author} {\bibfnamefont {W.}~\bibnamefont {Heil}},
  \bibinfo {author} {\bibfnamefont {S.}~\bibnamefont {Karpuk}}, \bibinfo
  {author} {\bibfnamefont {W.}~\bibnamefont {Kilian}}, \bibinfo {author}
  {\bibfnamefont {S.}~\bibnamefont {Knappe-Gr\"uneberg}}, \bibinfo {author}
  {\bibfnamefont {W.}~\bibnamefont {M\"uller}}, \bibinfo {author}
  {\bibfnamefont {U.}~\bibnamefont {Schmidt}}, \bibinfo {author} {\bibfnamefont
  {A.}~\bibnamefont {Schnabel}}, \bibinfo {author} {\bibfnamefont
  {F.}~\bibnamefont {Seifert}}, \bibinfo {author} {\bibfnamefont
  {Y.}~\bibnamefont {Sobolev}}, \ and\ \bibinfo {author} {\bibfnamefont
  {L.}~\bibnamefont {Trahms}},\ }\href {\doibase
  10.1103/PhysRevLett.111.100801} {\bibfield  {journal} {\bibinfo  {journal}
  {Phys. Rev. Lett.}\ }\textbf {\bibinfo {volume} {111}},\ \bibinfo {pages}
  {100801} (\bibinfo {year} {2013})}\BibitemShut {NoStop}%
\bibitem [{\citenamefont {Bloch}(1946)}]{Bloch:1946}%
  \BibitemOpen
  \bibfield  {author} {\bibinfo {author} {\bibfnamefont {F.}~\bibnamefont
  {Bloch}},\ }\href {\doibase 10.1103/PhysRev.70.460} {\bibfield  {journal}
  {\bibinfo  {journal} {Phys. Rev.}\ }\textbf {\bibinfo {volume} {70}},\
  \bibinfo {pages} {460} (\bibinfo {year} {1946})}\BibitemShut {NoStop}%
\bibitem [{\citenamefont {Chu}\ and\ \citenamefont {Peng}(2015)}]{Chu:2015bta}%
  \BibitemOpen
  \bibfield  {author} {\bibinfo {author} {\bibfnamefont {P.-H.}\ \bibnamefont
  {Chu}}\ and\ \bibinfo {author} {\bibfnamefont {J.-C.}\ \bibnamefont {Peng}},\
  }\href {\doibase 10.1016/j.nima.2015.05.062} {\bibfield  {journal} {\bibinfo
  {journal} {Nucl. Instrum. Meth.}\ }\textbf {\bibinfo {volume} {A795}},\
  \bibinfo {pages} {128} (\bibinfo {year} {2015})},\ \Eprint
  {http://arxiv.org/abs/1505.06406} {arXiv:1505.06406 [nucl-ex]} \BibitemShut
  {NoStop}%
\bibitem [{\citenamefont {Lee}\ and\ \citenamefont {Romalis}(2008)}]{Lee_2008}%
  \BibitemOpen
  \bibfield  {author} {\bibinfo {author} {\bibfnamefont {S.-K.}\ \bibnamefont
  {Lee}}\ and\ \bibinfo {author} {\bibfnamefont {M.~V.}\ \bibnamefont
  {Romalis}},\ }\href {\doibase 10.1063/1.2885711} {\bibfield  {journal}
  {\bibinfo  {journal} {Journal of Applied Physics}\ }\textbf {\bibinfo
  {volume} {103}},\ \bibinfo {pages} {084904} (\bibinfo {year}
  {2008})}\BibitemShut {NoStop}%
\bibitem [{\citenamefont {Wineland}\ \emph {et~al.}(1991)\citenamefont
  {Wineland}, \citenamefont {Bollinger}, \citenamefont {Heinzen}, \citenamefont
  {Itano},\ and\ \citenamefont {Raizen}}]{Wineland:1991}%
  \BibitemOpen
  \bibfield  {author} {\bibinfo {author} {\bibfnamefont {D.~J.}\ \bibnamefont
  {Wineland}}, \bibinfo {author} {\bibfnamefont {J.~J.}\ \bibnamefont
  {Bollinger}}, \bibinfo {author} {\bibfnamefont {D.~J.}\ \bibnamefont
  {Heinzen}}, \bibinfo {author} {\bibfnamefont {W.~M.}\ \bibnamefont {Itano}},
  \ and\ \bibinfo {author} {\bibfnamefont {M.~G.}\ \bibnamefont {Raizen}},\
  }\href {\doibase 10.1103/PhysRevLett.67.1735} {\bibfield  {journal} {\bibinfo
   {journal} {Phys. Rev. Lett.}\ }\textbf {\bibinfo {volume} {67}},\ \bibinfo
  {pages} {1735} (\bibinfo {year} {1991})}\BibitemShut {NoStop}%
\bibitem [{\citenamefont {Venema}\ \emph {et~al.}(1992)\citenamefont {Venema},
  \citenamefont {Majumder}, \citenamefont {Lamoreaux}, \citenamefont {Heckel},\
  and\ \citenamefont {Fortson}}]{Venema:1992}%
  \BibitemOpen
  \bibfield  {author} {\bibinfo {author} {\bibfnamefont {B.~J.}\ \bibnamefont
  {Venema}}, \bibinfo {author} {\bibfnamefont {P.~K.}\ \bibnamefont
  {Majumder}}, \bibinfo {author} {\bibfnamefont {S.~K.}\ \bibnamefont
  {Lamoreaux}}, \bibinfo {author} {\bibfnamefont {B.~R.}\ \bibnamefont
  {Heckel}}, \ and\ \bibinfo {author} {\bibfnamefont {E.~N.}\ \bibnamefont
  {Fortson}},\ }\href {\doibase 10.1103/PhysRevLett.68.135} {\bibfield
  {journal} {\bibinfo  {journal} {Phys. Rev. Lett.}\ }\textbf {\bibinfo
  {volume} {68}},\ \bibinfo {pages} {135} (\bibinfo {year} {1992})}\BibitemShut
  {NoStop}%
\bibitem [{\citenamefont {Youdin}\ \emph {et~al.}(1996)\citenamefont {Youdin},
  \citenamefont {Krause}, \citenamefont {Jagannathan}, \citenamefont {Hunter},\
  and\ \citenamefont {Lamoreaux}}]{Youdin:1996}%
  \BibitemOpen
  \bibfield  {author} {\bibinfo {author} {\bibfnamefont {A.~N.}\ \bibnamefont
  {Youdin}}, \bibinfo {author} {\bibfnamefont {D.}~\bibnamefont {Krause},
  \bibfnamefont {Jr.}}, \bibinfo {author} {\bibfnamefont {K.}~\bibnamefont
  {Jagannathan}}, \bibinfo {author} {\bibfnamefont {L.~R.}\ \bibnamefont
  {Hunter}}, \ and\ \bibinfo {author} {\bibfnamefont {S.~K.}\ \bibnamefont
  {Lamoreaux}},\ }\href {\doibase 10.1103/PhysRevLett.77.2170} {\bibfield
  {journal} {\bibinfo  {journal} {Phys. Rev. Lett.}\ }\textbf {\bibinfo
  {volume} {77}},\ \bibinfo {pages} {2170} (\bibinfo {year}
  {1996})}\BibitemShut {NoStop}%
\bibitem [{\citenamefont {Petukhov}\ \emph {et~al.}(2010)\citenamefont
  {Petukhov}, \citenamefont {Pignol}, \citenamefont {Jullien},\ and\
  \citenamefont {Andersen}}]{Petukhov:2010}%
  \BibitemOpen
  \bibfield  {author} {\bibinfo {author} {\bibfnamefont {A.~K.}\ \bibnamefont
  {Petukhov}}, \bibinfo {author} {\bibfnamefont {G.}~\bibnamefont {Pignol}},
  \bibinfo {author} {\bibfnamefont {D.}~\bibnamefont {Jullien}}, \ and\
  \bibinfo {author} {\bibfnamefont {K.~H.}\ \bibnamefont {Andersen}},\ }\href
  {\doibase 10.1103/PhysRevLett.105.170401} {\bibfield  {journal} {\bibinfo
  {journal} {Phys. Rev. Lett.}\ }\textbf {\bibinfo {volume} {105}},\ \bibinfo
  {pages} {170401} (\bibinfo {year} {2010})}\BibitemShut {NoStop}%
\bibitem [{\citenamefont {Bulatowicz}\ \emph {et~al.}(2013)\citenamefont
  {Bulatowicz}, \citenamefont {Griffith}, \citenamefont {Larsen}, \citenamefont
  {Mirijanian}, \citenamefont {Fu}, \citenamefont {Smith}, \citenamefont
  {Snow}, \citenamefont {Yan},\ and\ \citenamefont {Walker}}]{Bulatowicz:2013}%
  \BibitemOpen
  \bibfield  {author} {\bibinfo {author} {\bibfnamefont {M.}~\bibnamefont
  {Bulatowicz}}, \bibinfo {author} {\bibfnamefont {R.}~\bibnamefont
  {Griffith}}, \bibinfo {author} {\bibfnamefont {M.}~\bibnamefont {Larsen}},
  \bibinfo {author} {\bibfnamefont {J.}~\bibnamefont {Mirijanian}}, \bibinfo
  {author} {\bibfnamefont {C.~B.}\ \bibnamefont {Fu}}, \bibinfo {author}
  {\bibfnamefont {E.}~\bibnamefont {Smith}}, \bibinfo {author} {\bibfnamefont
  {W.~M.}\ \bibnamefont {Snow}}, \bibinfo {author} {\bibfnamefont
  {H.}~\bibnamefont {Yan}}, \ and\ \bibinfo {author} {\bibfnamefont {T.~G.}\
  \bibnamefont {Walker}},\ }\href {\doibase 10.1103/PhysRevLett.111.102001}
  {\bibfield  {journal} {\bibinfo  {journal} {Phys. Rev. Lett.}\ }\textbf
  {\bibinfo {volume} {111}},\ \bibinfo {pages} {102001} (\bibinfo {year}
  {2013})}\BibitemShut {NoStop}%
\bibitem [{\citenamefont {Guigue}\ \emph {et~al.}(2015)\citenamefont {Guigue},
  \citenamefont {Jullien}, \citenamefont {Petukhov},\ and\ \citenamefont
  {Pignol}}]{Guigue:2015fyt}%
  \BibitemOpen
  \bibfield  {author} {\bibinfo {author} {\bibfnamefont {M.}~\bibnamefont
  {Guigue}}, \bibinfo {author} {\bibfnamefont {D.}~\bibnamefont {Jullien}},
  \bibinfo {author} {\bibfnamefont {A.~K.}\ \bibnamefont {Petukhov}}, \ and\
  \bibinfo {author} {\bibfnamefont {G.}~\bibnamefont {Pignol}},\ }\href
  {\doibase 10.1103/PhysRevD.92.114001} {\bibfield  {journal} {\bibinfo
  {journal} {Phys. Rev.}\ }\textbf {\bibinfo {volume} {D92}},\ \bibinfo {pages}
  {114001} (\bibinfo {year} {2015})}\BibitemShut {NoStop}%
\bibitem [{\citenamefont {Lee}\ \emph {et~al.}(2018)\citenamefont {Lee},
  \citenamefont {Almasi},\ and\ \citenamefont {Romalis}}]{Lee:2018vaq}%
  \BibitemOpen
  \bibfield  {author} {\bibinfo {author} {\bibfnamefont {J.}~\bibnamefont
  {Lee}}, \bibinfo {author} {\bibfnamefont {A.}~\bibnamefont {Almasi}}, \ and\
  \bibinfo {author} {\bibfnamefont {M.}~\bibnamefont {Romalis}},\ }\href
  {\doibase 10.1103/PhysRevLett.120.161801} {\bibfield  {journal} {\bibinfo
  {journal} {Phys. Rev. Lett.}\ }\textbf {\bibinfo {volume} {120}},\ \bibinfo
  {pages} {161801} (\bibinfo {year} {2018})},\ \Eprint
  {http://arxiv.org/abs/1801.02757} {arXiv:1801.02757 [hep-ex]} \BibitemShut
  {NoStop}%
\bibitem [{\citenamefont {Raffelt}(2012)}]{Raffelt:2012sp}%
  \BibitemOpen
  \bibfield  {author} {\bibinfo {author} {\bibfnamefont {G.}~\bibnamefont
  {Raffelt}},\ }\href {\doibase 10.1103/PhysRevD.86.015001} {\bibfield
  {journal} {\bibinfo  {journal} {Phys. Rev.}\ }\textbf {\bibinfo {volume}
  {D86}},\ \bibinfo {pages} {015001} (\bibinfo {year} {2012})}\BibitemShut
  {NoStop}%
\end{thebibliography}%
\end{document}